\documentclass[]{interact}

\usepackage{amsmath}
\usepackage{epstopdf}
\usepackage[caption=false]{subfig}

\usepackage[numbers,sort&compress]{natbib}
\bibpunct[, ]{[}{]}{,}{n}{,}{,}
\makeatletter
\def\NAT@def@citea{\def\@citea{\NAT@separator}}
\makeatother

\theoremstyle{plain}

\theoremstyle{definition}

\theoremstyle{remark}

\begin{document}


\title{Applications of the Peach-Koehler Force in Liquid Crystals}

\author{
\name{Cheng Long\textsuperscript{a} and Jonathan V. Selinger*\textsuperscript{a}\thanks{CONTACT Jonathan V. Selinger. Email: jselinge@kent.edu}}
\affil{\textsuperscript{a}Department of Physics, Advanced Material and Liquid Crystal Institute, Kent State University, Kent, OH 44242, USA}
}

\maketitle

\begin{abstract}
In solids, external stress induces the Peach-Koehler force, which drives dislocations to move. Similarly, in liquid crystals, an external angular stress creates an analogous force, which drives disclinations to move. In this work, we develop a method to calculate the relevant angular stress either analytically or numerically, and hence to determine the force on a disclination. We demonstrate this method by applying the Peach-Koehler force theory to four problems: (a)~Single disclination in a liquid crystal cell between two uniform in-plane alignments perpendicular to each other. (b)~Array of disclinations in a liquid crystal cell with patterned substrates. (c)~Pair of disclinations in a long capillary tube with homeotropic anchoring. (d)~Radial hedgehog or disclination loop inside a sphere with homeotropic anchoring, and its response to an applied magnetic field. In all of these problems, the Peach-Koehler force theory predicts the equilibrium defect structure, and the predictions are consistent with the results of minimizing the total free energy.
\end{abstract}


\section{Introduction}

Topological defects are a manifestation of broken symmetries, and have structures topologically distinct from the uniform ground-state manifold determined by the Hamiltonian of a continuum field, because they cannot be continuously transformed into a uniform ground state. For physical systems with discrete translational symmetry, like periodic crystals, topological defects can take the form of dislocations in the displacement field. By comparison, for systems with uniaxial symmetry in the orientation distribution, like liquid crystals, topological defects are disclinations in the orientation field. Not only are topological defects important for fundamental physics, like configurations in the early universe~\cite{brandenberger1994topological,durrer2002cosmic} and phase transitions in superfluid helium and xy-models~\cite{kosterlitz1973order,agnolet1989kosterlitz,chaikin1995principles}, they also play an major role in determining properties of real materials. For example, dislocations mediate plastic deformation and melting in solids~\cite{frank1950multiplication,halperin1978theory}. In addition, researchers have found that topological defects can be exploited to understand and control living systems in active matter~\cite{peng2016command,genkin2017topological,zhang2021spatiotemporal,shankar2022topological,ardavseva2022topological}. Studying the dynamics of topological defects is essential for understanding this range of phenomena.

In crystalline solids, a dislocation is a topological line defect where the periodic translational symmetry is disrupted. Its properties are characterized by a Burgers vector $\boldsymbol{b}$, which indicates the local distortion of the displacement field, as well as a local unit tangent vector $\hat{\boldsymbol{t}}$. When a crystalline solid is subjected to an external stress $\boldsymbol{\sigma}$, dislocations inside the solid experience a local Peach-Koehler force per length of $\boldsymbol{F}_{\mathrm{PK}} = (\boldsymbol{\sigma}\cdot \boldsymbol{b})\times \hat{\boldsymbol{t}}$~\cite{peach1950forces,anderson2017theory}. This force is responsible for plastic deformation of metals due to the multiplication of pinned dislocations.

In nematic liquid crystals, a disclination is a line defect where the uniaxial symmetry is destroyed and the director is not well-defined. As one moves in a loop around the disclination, the director rotates by an angle of $\pi$, returning to the starting orientation thanks to the uniaxial symmetry of the director outside the defect core region. The local axis of rotation is denoted by $\hat{\boldsymbol{\Omega}}$~\cite{de1995physics}. The vector $\pi \hat{\boldsymbol{\Omega}}$ can be regarded as an effective Burgers vector, which combines information about the rotation angle and axis. Theoretical research has demonstrated that a disclination experiences an analogous Peach-Koehler force when it is subjected to an external angular stress~\cite{kleman1983points,eshelby1980force,kawasaki1985gauge,rey1990defect}. In previous work, our group showed that the analogous force per length has the form
\begin{equation}\label{sec1:FPK}
  \boldsymbol{F}_{\mathrm{PK}}= \left( \pi \hat{\boldsymbol{\Omega}} \cdot \boldsymbol{\sigma}^{\mathrm{ext}} \right) \times \hat{\boldsymbol{t}},
\end{equation}
where $\boldsymbol{\sigma}^{\mathrm{ext}}$ represents the external angular stress acting on the disclination in the one-constant approximation~\cite{long2021geometry}. In recent years, the concept of Peach-Koehler force has been successfully applied to solve a few simple theoretical models for the dynamics of disclinations~\cite{schimming2022singularity,schimming2023kinematics,long2022frank}. However, in most practical circumstances, external angular stress can be induced by complicated boundary conditions, interactions with other disclinations, and self-interaction for a curved disclination, and hence it is difficult to calculate.

The purpose of this article is to develop analytic and numerical methods to use the Peach-Koehler force theory for liquid crystals, and to demonstrate these methods for several problems of current interest. In general, we show that the external angular stress $\boldsymbol{\sigma}^{\mathrm{ext}}$ acting on a dislination is equal to the total stress field $\sigma_{ij}^{\mathrm{total}} = K \epsilon_{ikl}n_k\partial_j n_l$, minus the stress field induced by a free straight disclination with the same $\hat{\boldsymbol{\Omega}}$, evaluated at the disclination itself. This difference turns out to be the zeroth-order term in the power series of $\sigma_{ij}^{\mathrm{total}}$ expanded around the disclination. Using this theory, we investigate the equilibrium state for several liquid crystal systems with disclinations. In Section~2, we consider a single twist disclination line inside a nematic liquid crystal cell between two infinite parallel plates with uniform alignments, and use the Peach-Koehler force to determine the equilibrium position of the disclination line. In Section~3, we discuss a row of parallel twist disclination lines in a liquid crystal cell between two infinite parallel plates, where one plate has a periodic alignment pattern. We calculate the Peach-Koehler forces acting on the twist disclinations, to see the effect of the periodic alignment pattern. In Section~4, we study two parallel $+1/2$ disclination lines in a long cylindrical capillary tube with homeotropic anchoring. Here, the Peach-Koehler force theory predicts analytically the optimum distance between the two disclination lines. In Section~5, we apply the Peach-Koehler force theory to a $+1/2$ wedge disclination loop in a spherical droplet with homeotropic anchoring, and discuss the equilibrium size of the disclination loop with or without an applied magnetic field.

\section{Single twist disclination line in a liquid crystal cell}

\begin{figure}
\centering
\includegraphics[width=0.8\textwidth]{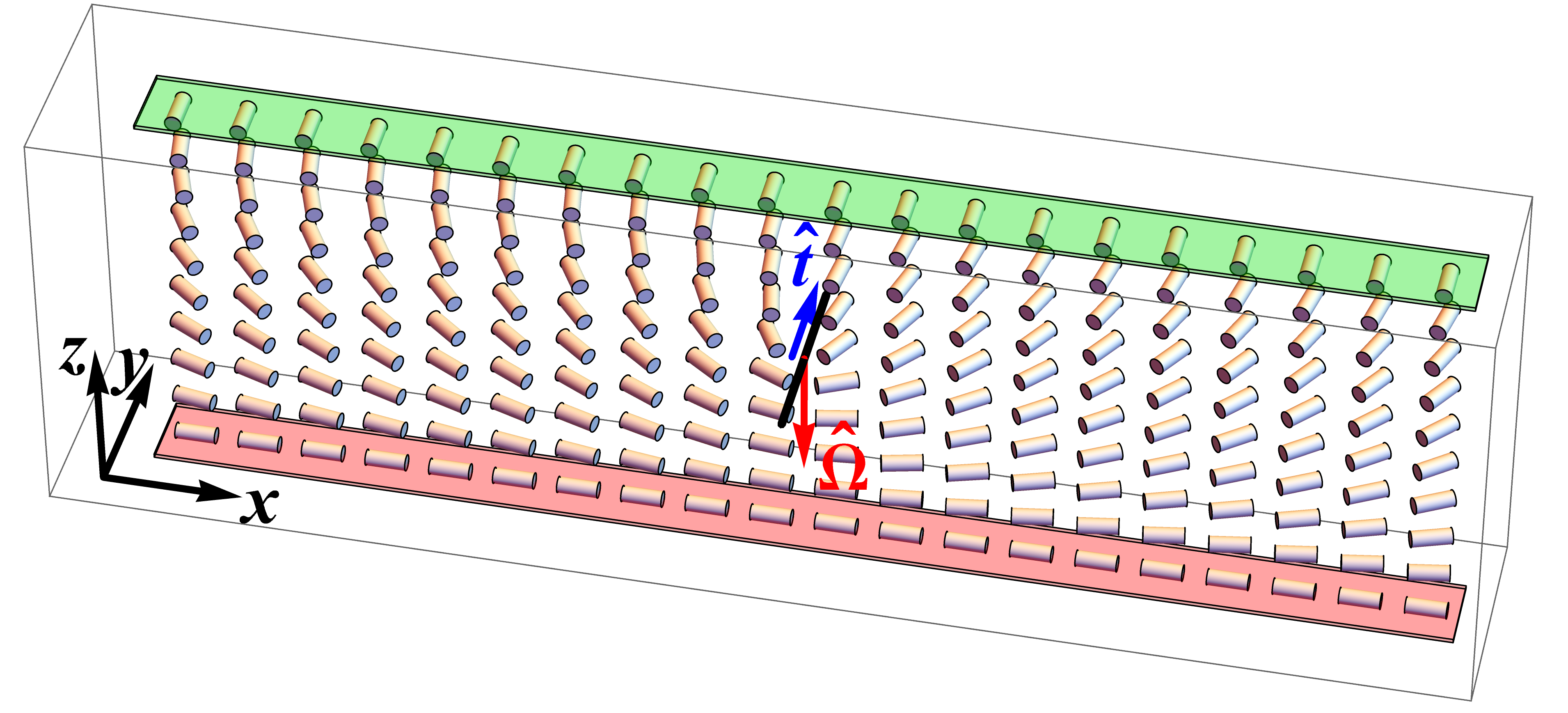}
\caption{Schematic illustration of a single straight twist disclination line in a nematic liquid crystal cell between two parallel substrates. The thickness of the cell is $L_z$. The surface anchoring on the top substrate (green) requires the director aligned with the $y$-axis, while that on the bottom substrate (red) requires the director aligned with the $x$-axis. The director field forms regions of right- and left-handed twist. Between these regions, there is a twist disclination line, shown in black, parallel to the $y$ direction.} \label{fig1:singledisclination}
\end{figure}

As a simple example, we first consider the equilibrium position of a single straight twist disclination line in a nematic liquid crystal cell between two parallel substrates, illustrated in Figure~\ref{fig1:singledisclination}. On each substrate, there is uniform strong anchoring, and the anchoring directions on the two substrates are perpendicular to each other. Between the substrates, the director can twist in a right- or left-handed helix, and these two twist directions have equal free energy for an achiral liquid crystal. Twist disclination lines form at the boundaries between regions of different twist directions. In this section, we discuss an isolated twist disclination line extending in the $y$ direction.

Because of the planar surface anchoring, the director inside the liquid crystal cell is confined in the $x$-$y$ plane parallel to the substrates and can be described by
\begin{equation}\label{sec2:director}
  \hat{\boldsymbol{n}} = \left( \cos{\phi}, \sin{\phi}, 0 \right).
\end{equation}
On a loop around the disclination, the director rotates around the $z$-axis. If we define the tangent vector of the straight disclination line as $\hat{\boldsymbol{t}}=(0,1,0)$, then the rotation axis of the director around the disclination in Figure~\ref{fig1:singledisclination} is $\hat{\boldsymbol{\Omega}}=(0,0,-1)$. Because the system has continuous translational symmetry along the $y$-axis, the director field becomes $\hat{\boldsymbol{n}}(x,z)$, independent of $y$. For a nematic liquid crystal with equal Frank constants, the total free energy is
\begin{equation}\label{sec2:freeenergy}
  \frac{F}{L_y} = \int \frac{1}{2} K (\partial_i n_j)(\partial_i n_j) \  \mathrm{d}x \mathrm{d}z 
  = \int \frac{1}{2} K |\boldsymbol{\nabla}\phi|^2 \  \mathrm{d}x \mathrm{d}z,
\end{equation}
using the Einstein summation convention. In this expression, $L_y$ is the dimension of the liquid crystal cell in the $y$ direction, and $K$ is the Frank elastic constant. To find the stable director configuration around the disclination, we have to solve the Euler-Lagrange equation derived from the total free energy, which is
\begin{equation}\label{sec2:pdeforphi}
  K \nabla^2 \phi = 0.
\end{equation}
Without loss of generality, we suppose the disclination is located at $x=0$ and $z=d$, and we define the branch cut in $\phi$ straight up to the top substrate at $z=L_z/2$. The boundary conditions then become $\phi(x,-L_z/2)=0$, $\phi(x,L_z/2)=-\pi/2$ for $x<0$, and $\phi(x,L_z/2)=\pi/2$ for $x>0$.

This model was extensively studied by Wang et al.~\cite{wang2017artificial}, who solved the Euler-Lagrange equation~(\ref{sec2:pdeforphi}) using the method of conformal mapping. Their solution is
\begin{eqnarray}\label{sec2:solutionforphi}
  \phi \left(x,z\right) = \frac{1}{2} \left[ \arctan{\left( \frac{\displaystyle \sin{\frac{\pi z}{L_z}}\cosh{\frac{\pi x}{L_z}}-\sin{\frac{\pi d}{L_z}}}{\displaystyle \cos{\frac{\pi z}{L_z}}\sinh{\frac{\pi x}{L_z}}} \right)} + \frac{\pi}{2}\mathrm{sgn}\left(x\right) \right],
\end{eqnarray}
where $\mathrm{sgn}$ returns the sign of $x$. This solution applies for any value of $d$, and hence it does not determine the equilibrium value of $d$. Based on symmetry, it is clear that the equilibrium position should be $d=0$, halfway between the substrates at $z=\pm L_z/2$. 
To demonstrate this point explicitly, Wang et al.\ put the solution (\ref{sec2:solutionforphi}) back into the free energy (\ref{sec2:freeenergy}), and integrate to find the total free energy as a function of $d$. This calculation shows that $d=0$ gives the minimum free energy, and it shows the restoring force if the disclination is displaced to a different height $d\neq0$. Here, we offer an alternative approach from the perspective of the Peach-Koehler force.

In this system, the only force acting on the disclination is the Peach-Koehler force, which is caused by the boundary constraints. To calculate the Peach-Koehler force, we need the external angular stress acting on the disclination. From our previous paper~\cite{long2021geometry}, the total angular stress is given by
\begin{equation}\label{sec2:sigmatotal}
  \sigma^{\mathrm{total}}_{ij} = K \epsilon_{ikl} n_k \partial_j n_l.
\end{equation}
In this expression, we insert the director field from Eq.~(\ref{sec2:director}), and then the angle $\phi(x,z)$ from Eq.~(\ref{sec2:solutionforphi}). Because we are interested in the angular stress close to the disclination, we change variables from $(x,z)$ to polar coordinates $(\rho,\varphi)$ centered at the disclination
\begin{equation}\label{sec2:coordinatetransform}
  x = \rho \cos{\varphi},\quad
  z = d + \rho \sin{\varphi}.
\end{equation}
We then expand $\boldsymbol{\sigma}^{\mathrm{total}}$ as a power series around the disclination point $\rho=0$, and obtain
\begin{equation}
  \boldsymbol{\sigma}^{\mathrm{total}} = K
  \begin{pmatrix}
    0 & 0 & 0 \\
    0 & 0 & 0 \\
    - \dfrac{\sin{\varphi}}{2\rho} + \dfrac{ \pi \tan{\frac{\pi d}{L_z}}}{4 L_z} & 0 & \dfrac{\cos{\varphi}}{2\rho}
  \end{pmatrix} + O(\rho).
\end{equation}
Because the director changes rapidly in space close to the disclination core, the total angular stress diverges as $1/\rho$ approaching the disclination core. Since the director field is not symmetric about the tangent vector of the disclination, the total angular stress also depends on $\varphi$.

To determine the \emph{external} angular stress due to the boundary conditions, we must subtract the angular stress of a \emph{free} disclination at $x=0$ and $z=d$. For a free disclination without any boundary constraints, the angle $\phi(x,z)$ is given by
\begin{equation}\label{sec2:phifree}
  \phi_{\mathrm{free}}\left(x,z\right) = \frac{1}{2}\left( \arctan{\frac{z-d}{x}} + \frac{\pi}{2} \mathrm{sgn}(x) \right).
\end{equation}
(Here, the $\mathrm{sgn}(x)$ term makes the branch cut go vertically upward, consistent with the previous case.) The angular stress field induced by this free twist disclination is then
\begin{equation}\label{sec2:stressforfreedisclination}
  \boldsymbol{\sigma}^{\mathrm{free}} = K
  \begin{pmatrix}
    0 & 0 & 0 \\
    0 & 0 & 0 \\
    -\dfrac{\sin{\varphi}}{2\rho} & 0 & \dfrac{\cos{\varphi}}{2\rho}
  \end{pmatrix}.
\end{equation}
in the polar coordinates $(\rho,\varphi)$. We can see that $\boldsymbol{\sigma}^{\mathrm{free}}$ is proportional to $K/\rho$. Indeed, for a liquid crystal with a single Frank constant $K$, the angular stress of a \emph{free} disclination should always be proportional to $K/\rho$, because $K$ is the only parameter with dimensions of force, and $\rho$ is the only variable with dimensions of length.

Now that we have $\boldsymbol{\sigma}^{\mathrm{free}}$, we can determine the \emph{external} angular stress acting on the disclination by subtracting $\boldsymbol{\sigma}^{\mathrm{ext}} = \boldsymbol{\sigma}^{\mathrm{total}} - \boldsymbol{\sigma}^{\mathrm{free}}$. In this difference, the $1/\rho$ term in $\boldsymbol{\sigma}^{\mathrm{total}}$ is cancelled by $\boldsymbol{\sigma}^{\mathrm{free}}$, and only the zeroth- and higher-order terms remain. Because we want the external angular stress at the disclination, we take the limit $\rho \to 0$. In that limit, all of the higher-order terms vanish, and only the zeroth-order term remains. Hence, the external angular stress acting on the disclination becomes
\begin{equation}
  \boldsymbol{\sigma}^{\mathrm{ext}} = \lim_{\rho \to 0} \left( \boldsymbol{\sigma}^{\mathrm{total}} - \boldsymbol{\sigma}^{\mathrm{free}} \right) = K
  \begin{pmatrix}
    0 & 0 & 0 \\
    0 & 0 & 0 \\
    \dfrac{ \pi \tan{\frac{\pi d}{L_z}}}{4 L_z} & 0 & 0
  \end{pmatrix}.
\end{equation}
We now put this expression for $\boldsymbol{\sigma}^{\mathrm{ext}}$, along with the unit vectors $\hat{\boldsymbol{\Omega}}$ and $\hat{\boldsymbol{t}}$, into Eq.~(\ref{sec1:FPK}) to obtain the Peach-Koehler force per length acting on the disclination line,
\begin{equation}\label{sec2:pkforce}
  \boldsymbol{F}_{\mathrm{PK}} = \left( 0, 0, -\dfrac{\pi^2 K}{4 L_z}\tan{\dfrac{\pi d}{L_z}} \right).
\end{equation}
This expression for the Peach-Koehler force has no $x$ or $y$ component, because of the symmetry of the liquid crystal cell. The $z$ component indicates how the twist disclination interacts with the top and the bottom substrates. When $d=0$, the Peach-Koehler force is zero, and the disclination line is at its equilibrium position. When $d>0$, the force is in the negative $z$ direction, pushing the disclination back to its equilibrium. If $d$ approaches $L_z/2$, the force diverges due to the strong repulsion from the top substrate. Likewise, when $d<0$, the Peach-Koehler force has the same behavior but with the opposite sign. We can see that the Peach-Koehler force is proportional to $K/L_z$. If $K$ increases, the Peach-Koehler force becomes larger because the liquid crystal has stronger elasticity. If $L_z$ increases, the force becomes weaker because the boundaries are farther from the disclination.

Our expression for the Peach-Koehler force in Eq.~(\ref{sec2:pkforce}) is precisely consistent with the result derived from calculating the total free energy of the liquid crystal cell in Ref.~\cite{wang2017artificial}. Hence, our Peach-Koehler theory provides an alternative way to find the equilibrium position of the disclination line without integrating the free energy density over the entire liquid crystal cell.

In most liquid crystal systems, analytic solutions for director configurations satisfying the Euler-Lagrange equations are not available, and one must rely on numerical solutions to analyze the director fields. Here, in addition to our analytical approach above, we provide a numerical approach to calculate the Peach-Koehler force acting on the disclination and find its equilibrium position. Since the director field close to a disclination changes rapidly as a function of position, the information about the external angular stress acting on a disclination in a numerical solution is blurred by numerical uncertainty created by the rapid variation of the director. To deal with this problem, we decompose the angle field $\phi(x,z)$ into a singular part (which is known analytically) and a nonsingular part (which must be found numerically),
\begin{equation}\label{sec2:phidecomposed}
  \phi(x,z) = \phi_{\mathrm{free}}(x,z) + \phi_{\mathrm{ns}}(x,z),
\end{equation}
with $\phi_{\mathrm{free}}(x,z)$ given by Eq.~(\ref{sec2:phifree}). In this expression, the first part represents the angle field for a free twist disclination at $x=0$ and $z=d$. It has a singularity where the director gradient diverges. The second part is a nonsingular field, required by the boundary conditions, which is added to the angle field for the free twist disclination.

We insert the decomposed $\phi(x,z)$ into the director field $\hat{\boldsymbol{n}}$ of Eq.~(\ref{sec2:director}), and put that director field into total angular stress tensor $\boldsymbol{\sigma}^{\mathrm{total}}$ of Eq.~(\ref{sec2:sigmatotal}). We then subtract off the angular stress of the free twist disclination in Eq.~(\ref{sec2:stressforfreedisclination}). The remaining \emph{external} angular stress tensor is then given by
\begin{equation}
  \boldsymbol{\sigma}^{\mathrm{ext}} = K \lim_{\substack{x \to 0 \\ z \to d}}
  \begin{pmatrix}
    0 & 0 & 0 \\
    0 & 0 & 0 \\
    \partial_x \phi_{\mathrm{ns}} & 0 & \partial_z \phi_{\mathrm{ns}}
  \end{pmatrix}.
\end{equation}
Using this result, the Peach-Koehler force acting on the disclination becomes
\begin{equation}
  \boldsymbol{F}_{\mathrm{PK}} = \pi K \lim_{\substack{x \to 0 \\ z \to d}} \left( \partial_z \phi_{\mathrm{ns}}, 0, -\partial_x \phi_{\mathrm{ns}} \right).
\end{equation}
As expected, the Peach-Koehler force acting on the disclination depends only on $\phi_{\mathrm{ns}}(x,z)$. The derivative of $\partial_x\phi_{\mathrm{ns}}$ determines the $z$-component of the Peach-Koehler force. Likewise, the derivative $\partial_z\phi_{\mathrm{ns}}$ could give rise to a nonzero $x$-component of the Peach-Koehler force, but we will see later this derivative vanishes because of the symmetry of the director field.

\begin{figure}
\centering
\includegraphics[width=1\textwidth]{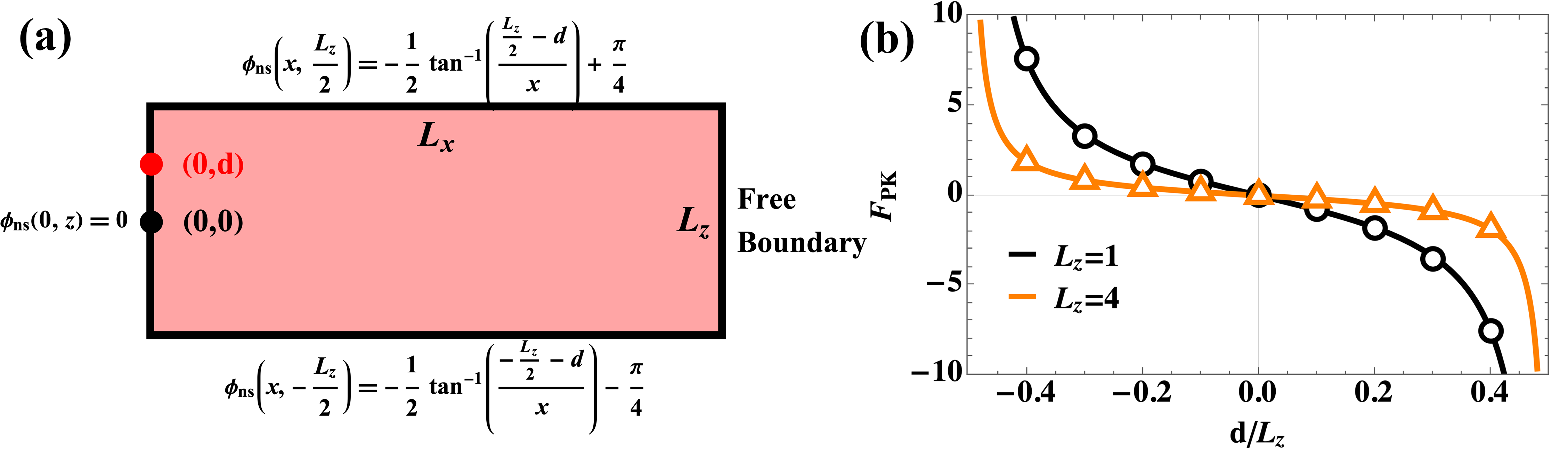}
\caption{(a)~Illustration of the domain for the nonsingular field $\phi_{\mathrm{ns}}(x,z)$. The black dot on the left boundary is the origin of the coordinate system $(0,0)$, and the red dot represents the disclination at $(0,d)$. We fix the length $L_x=10$, and vary $d$ and $L_z$. (b)~Peach-Koehler force calculated numerically as a function of $d/L_z$. Black circles are numerical results for $L_z=1$, and orange triangles are numerical results for $L_z=4$. Solid lines are plotted based on the analytic form in Eq.~(\ref{sec2:pkforce}).} \label{fig2:numericalapproach}
\end{figure}

Now let us discuss a numerical method for calculating $\phi_{\mathrm{ns}}(x,z)$. The fields $\phi(x,z)$ and $\phi_{\mathrm{free}}(x,z)$ both obey the Euler-Lagrange equation~(\ref{sec2:pdeforphi}), which is Laplace's equation, and it is a \emph{linear} partial differential equation. Hence, the difference $\phi_{\mathrm{ns}}(x,z)$ must also satisfy the same partial differential equation. Taking account of the mirror symmetry of the director field about the $y$-$z$ plane crossing the disclination, we solve this equation for $\phi_{\mathrm{ns}}(x,z)$ in a rectangular domain of $L_x \times L_z$ on the right side of the mirror plane, illustrated in Figure~\ref{fig2:numericalapproach}(a). The disclination is located on the left side of the simulation box at $(0,d)$.

The boundary conditions for $\phi_{\mathrm{ns}}(x,z)$ are just the difference between the boundary conditions for $\phi(x,z)$ and the known boundary values of $\phi_{\mathrm{free}}(x,z)$. On the top and bottom boundaries, $\phi_{\mathrm{ns}}(x,z)$ satisfies Dirichlet boundary conditions as shown in the figure. On the left boundary, $\phi(x,z)$ and $\phi_{\mathrm{free}}(x,z)$ are identical ($0$ below the disclination, $\pi/2$ above the disclination), and hence we have the Dirichlet boundary condition $\phi_{\mathrm{ns}}(0,z)=0$. (This boundary condition implies that $\partial_z\phi_{\mathrm{ns}}=0$ at the disclination, and hence the $x$-component of the Peach-Koehler force vanishes.) On the right side of the domain, we use a free boundary condition to reduce the boundary effect caused by finite $L_x$.

We solve the Euler-Lagrange equation for $\phi_{\mathrm{ns}}$, subject to those boundary conditions, using COMSOL multiphysics software. In the calculations, we vary the parameters $d$ and $L_z$, with constant $L_x=10$. Our numerical results are shown in Figure~\ref{fig2:numericalapproach}(b). When $L_z=1$, the Peach-Koehler force is zero in the middle of the cell at $d=0$. As the disclination moves from $d=0$ toward either of the uniform substrates, the Peach-Koehler force increases to push it back to the equilibrium position. The force diverges as $d\to\pm L_z/2$ due to strong repulsion from the substrates. When $L_z=4$, the equilibrium position of the disclination is still at $d=0$, but the Peach-Koehler force becomes weaker compared to $L_z=1$. This reduced force is reasonable because the the disclination is farther from the substrates at the same ratio $d/L_z$. Our numerical results are consistent with the Peach-Koehler force derived analytically in Eq.~(\ref{sec2:pkforce}).

\section{Array of twist disclination lines in a liquid crystal cell}

\begin{figure}
\centering
\includegraphics[width=1\textwidth]{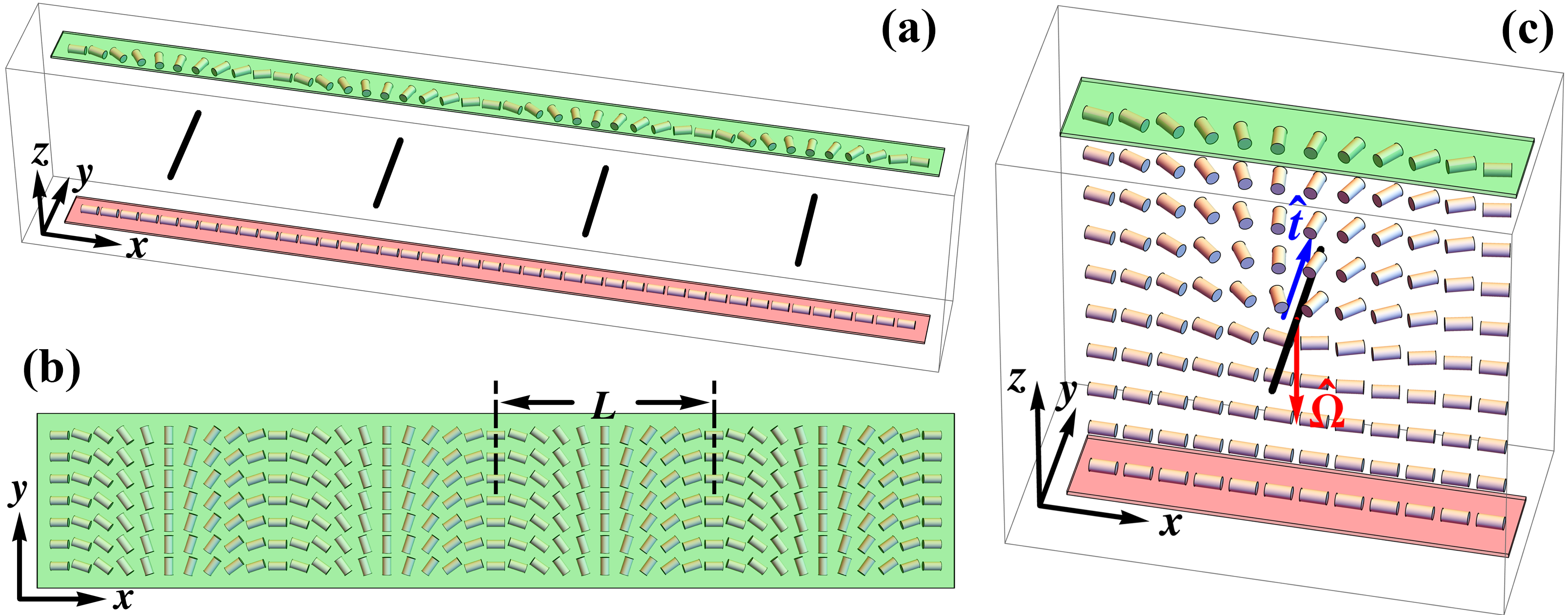}
\caption{(a)~Schematic illustration of a series of straight twist disclinations in a nematic liquid crystal cell between two parallel substrates. The thickness of the cell is $L_z$. The black lines parallel to the $y$-axis inside the cell represent the twist disclination lines. The surface anchoring on the bottom substrate (red) requires the director aligned with the $x$-axis. (b)~The surface anchoring on the top substrate (green) has a periodic pattern, with periodicity $L$ along the $x$ direction. (c) Director field inside a unit cell, with $x$ from $-L/2$ to $L/2$, and $z$ from $-L_z/2$ to $L_z/2$.} \label{fig3:arowofdisclinations}
\end{figure}

For a second example of the Peach-Koehler force, we are motivated by recent research interest in liquid crystal applications for smart soft coatings involving topological defects~\cite{babakhanova2020dynamically}. We consider a series of parallel straight twist disclination lines in a nematic liquid crystal cell between two substrates, as in Figure~\ref{fig3:arowofdisclinations}(a). The bottom substrate has uniform anchoring in the $x$-direction, and the top substrate has the periodic pattern shown in Figure~\ref{fig3:arowofdisclinations}(b):  Moving along the $x$-direction, the top anchoring orientation rotates around the $z$-axis continuously with a periodicity of $L$. This combination of substrates changes the relative orientation between bottom and top alternately between parallel and perpendicular. Twist disclination lines form at the locations where the bottom and top orientations are perpendicular to each other, as illustrated in the detailed view of Figure~\ref{fig3:arowofdisclinations}(c). In the Cartesian coordinates shown in the figure, the tangent vectors of the disclination lines can be defined as $\hat{\boldsymbol{t}}=(0,1,0)$, and the corresponding rotation axis of the director around the disclinations is $\hat{\boldsymbol{\Omega}}=(0,0,-1)$.

This model is similar in many ways to the model in the previous section. Because there is a continuous translational symmetry along the $y$-axis, the director field $\hat{\boldsymbol{n}}(x,z)$ does not depend on $y$. Because the anchoring on both substrates requires the director to be in the $x$-$y$ plane, the director field inside the liquid crystal cell can be described by $\hat{\boldsymbol{n}} = ( \cos{\phi}, \sin{\phi}, 0 )$, as in Eq.~(\ref{sec2:director}). The free energy is still given by Eq.~(\ref{sec2:freeenergy}), and the corresponding Euler-Lagrange equation by Eq.~(\ref{sec2:pdeforphi}).

Because of the periodicity, it is sufficient to calculate the equilibrium director field in a unit cell for $x$ from $-L/2$ to $L/2$, and $z$ from $-L_z/2$ to $L_z/2$. The solution can then be repeated periodically in $x$. The boundary conditions for this unit cell are $\phi(x,-L_z/2) = 0$ for the strong anchoring on the bottom substrate, $\phi(x,L_z/2) = -\pi [x/L-\mathrm{sgn}(x/L)/2]$ for the strong anchoring on the top substrate, and $\phi(-L/2,z)=\phi(L/2,z)$ on the periodic boundaries.

This model has been studied both experimentally and theoretically by Babakhanova et al.~\cite{babakhanova2020dynamically}.  On the theoretical side, they found a solution for $\phi(x,z)$ assuming the disclination line in the unit cell is fixed at $x=0$ and $z=d$. In our notation, their solution can be written as
\begin{eqnarray}\label{sec3:solutionforphi}
  \phi \left(x,z\right) = \frac{1}{2} \left[ \arctan{\left( \frac{\displaystyle \sin{\frac{\pi z}{L_z}}\cosh{\frac{\pi x}{L_z}}-\sin{\frac{\pi d}{L_z}}}{\displaystyle \cos{\frac{\pi z}{L_z}}\sinh{\frac{\pi x}{L_z}}} \right)} + \frac{\pi}{2}\mathrm{sgn}\left(x\right) \right] - \dfrac{\pi(z+L_z/2)}{L_z} \dfrac{x}{L}.
\end{eqnarray}
We should mention that this solution satisfies the Euler-Lagrange equation and the boundary conditions when $L \gg L_z$, but it slightly violates the periodic boundary condition otherwise. We will use it with the assumption that $L \gg L_z$.

Once again, as in the previous section, this solution applies for any disclination height $d$, and hence it does not determine the equilibrium value of $d$. Unlike the previous section, there is no symmetry to tell us the equilibrium height of the disclinations. For that reason, Ref.~\cite{babakhanova2020dynamically} found the equilibrium height by putting the solution back into the free energy, integrating numerically over the unit cell, and minimizing the total free energy numerically as a function of $d$. Here, we want to use the analytic solution directly to calculate the Peach-Koehler force, and then determine the equilibrium height where the Peach-Koehler force vanishes.

In principle, the Peach-Koehler force on a disclination in this model could be caused by interactions with the top and bottom surfaces, or by interactions with the other parallel disclinations. However, the interactions with the other disclinations cancel out because of mirror symmetry in the the $y$-$z$ plane. Hence, only interactions between the disclination and the two substrates contribute to the final Peach-Koehler force.

To calculate this force, we follow the same procedure as in the previous section. First, we combine the analytic solution of Eq.~(\ref{sec3:solutionforphi}) with the previous Eqs.~(\ref{sec2:director}) and~(\ref{sec2:sigmatotal}) to calculate the total angular stress tensor $\boldsymbol{\sigma}^{\mathrm{total}}$. Next, we change the coordinates from $(x,z)$ to polar coordinates $(\rho,\varphi)$ centered on the disclination, defined in Eq.~(\ref{sec2:coordinatetransform}). We expand the total angular stress as a power series about $\rho=0$, and obtain
\begin{equation}
  \boldsymbol{\sigma}^{\mathrm{total}} = K
  \begin{pmatrix}
    0 & 0 & 0 \\
    0 & 0 & 0 \\
    - \dfrac{\sin{\varphi}}{2\rho} + \dfrac{ \pi \tan{\frac{\pi d}{L_z}}}{4 L_z} - \dfrac{\pi}{L}\left( \dfrac{d}{L_z}+\dfrac{1}{2} \right) & 0 & \dfrac{\cos{\varphi}}{2\rho}
  \end{pmatrix} + O(\rho).
\end{equation}
We then subtract the angular stress tensor of a free twist disclination at $x=0$ and $z=d$, which is still given by Eq.~(\ref{sec2:stressforfreedisclination}), and take the limit of $\rho\to0$. From this calculation, the external stress acting on the disclination is found to be
\begin{equation}
  \boldsymbol{\sigma}^{\mathrm{ext}} = \lim_{\rho \to 0} \left( \boldsymbol{\sigma}^{\mathrm{total}} - \boldsymbol{\sigma}^{\mathrm{free}} \right) = K
  \begin{pmatrix}
    0 & 0 & 0 \\
    0 & 0 & 0 \\
    \dfrac{ \pi \tan{\frac{\pi d}{L_z}}}{4 L_z} - \dfrac{\pi}{L}\left( \dfrac{d}{L_z}+\dfrac{1}{2} \right) & 0 & 0
  \end{pmatrix}.
\end{equation}
By putting this expression for $\boldsymbol{\sigma}^{\mathrm{ext}}$, as well as the unit vectors $\hat{\boldsymbol{\Omega}}$ and $\hat{\boldsymbol{t}}$, into Eq.~(\ref{sec1:FPK}), we obtain the Peach-Koehler force per length acting on a disclination
\begin{equation}\label{sec3:pkforce}
  \boldsymbol{F}_{\mathrm{PK}} = \left( 0, 0, -\dfrac{\pi^2 K}{4 L_z}\tan{\dfrac{\pi d}{L_z}} + \dfrac{\pi^2 K}{L} \left( \dfrac{d}{L_z} + \dfrac{1}{2} \right) \right).
\end{equation}

As we can see in Eqn.~(\ref{sec3:pkforce}), the Peach-Koehler force has no $x$ or $y$ component because of the symmetries of the director field. The $z$ component comes from the interaction of the disclination with the uniform bottom substrate and patterned top substrate. The Peach-Koehler force has two terms. The first term is exactly the same as in the previous section, and is proportional to $K/L_z$. It dominates when the disclinations are close to the bottom or top substrate. The second term is a new effect of the periodicity in the top substrate, and is proportional to $K/L$. This second term always provides a positive force, which drives the disclinations closer to the patterned top substrate and farther from the uniform bottom substrate. Because of this term, the equilibrium position of the disclinations is not at $d=0$, but at some positive value of $d$.

\begin{figure}
\centering
\includegraphics[width=.55\textwidth]{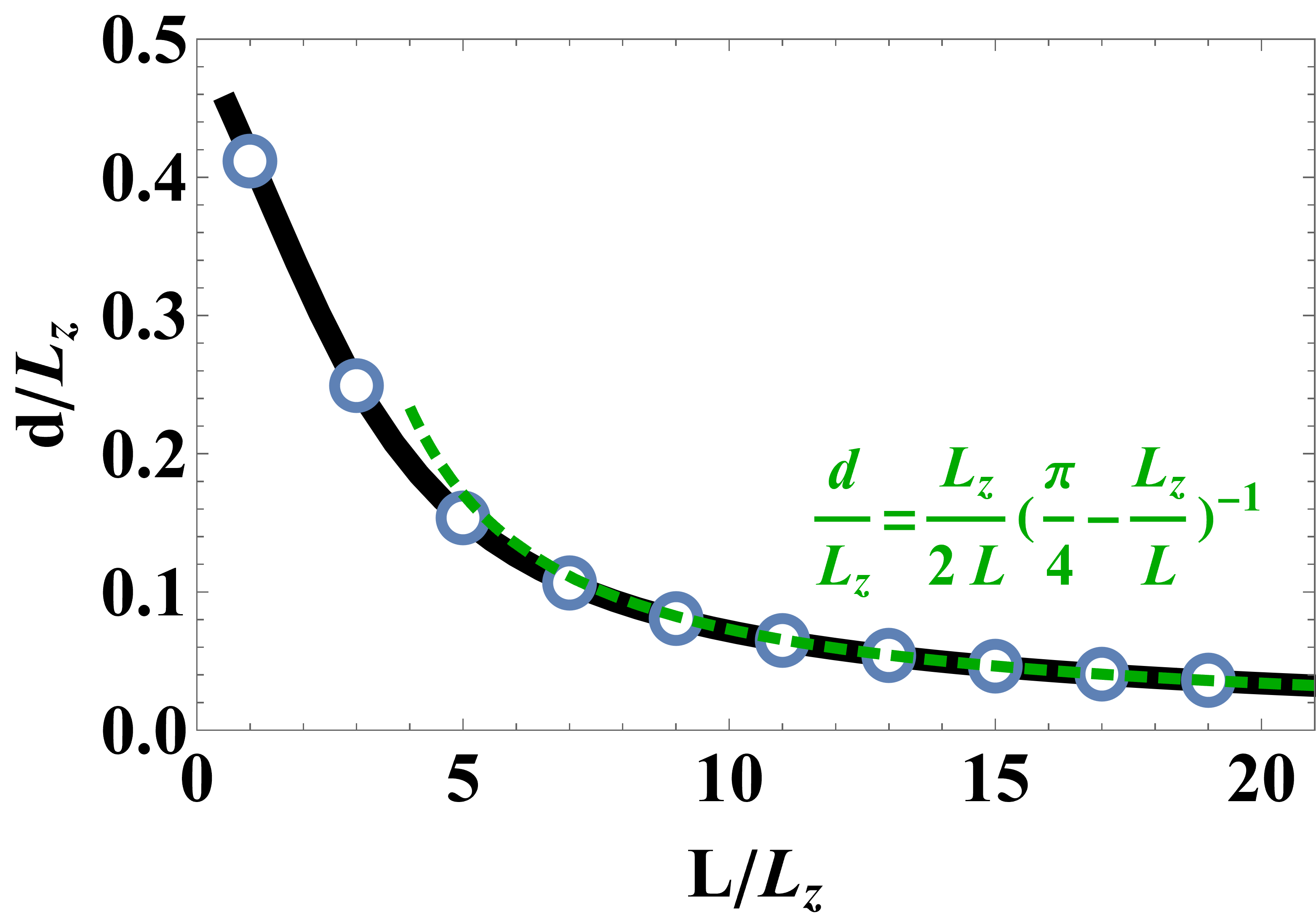}
\caption{Reduced equilibrium position $d/L_z$ of the disclinations, as a function of the ratio $L/L_z$. The black solid line shows the numerical results from solving $\boldsymbol{F}_{\mathrm{PK}}=0$, and the green dashed line represents the approximate analytic solution of Eq.~(\ref{sec3:dapproximation}) in the limit $L\gg L_z$. The blue circles are data points calculated by numerically minimizing the total free energy of the director field of Eq.~(\ref{sec3:solutionforphi}), integrated over the unit cell.} \label{fig4:equilibriumpos}
\end{figure}

To find the equilibrium position of the disclinations, we solve the equation $\boldsymbol{F}_{\mathrm{PK}}=0$ numerically for the reduced position $d/L_z$ as a function of the ratio $L/L_z$. Our numerical results are shown by the black line in Figure~\ref{fig4:equilibriumpos}. These results show that the reduced position $d/L_z$ decreases monotonically as a function of $L/L_z$. When the period of the top pattern is much greater than the thickness of the cell ($L \gg L_z$), the first term in Eq.~(\ref{sec3:pkforce}) dominates and the equilibrium value of $d$ approaches zero. In that limit, we can solve the equation $\boldsymbol{F}_{\mathrm{PK}}=0$ approximately as
\begin{equation}\label{sec3:dapproximation}
  \frac{d}{L_z} = \frac{L_z}{2L}\left( \frac{\pi}{4} - \frac{L_z}{L} \right)^{-1}.
\end{equation}
As a check on these results, we put the director field of Eq.~(\ref{sec3:solutionforphi}) into the free energy, integrate numerically to find the total free energy as a function of $d$, and minimize numerically to find the equilibrium value of $d$ (which is the same calculation as in Ref.~\cite{babakhanova2020dynamically}). The results are shown by the blue circles in Figure~\ref{fig4:equilibriumpos}, and they are consistent with the Peach-Koehler force theory.

\section{Two $+1/2$ wedge disclination lines in a cylinder}

When a nematic liquid crystal is confined inside a cylinder with homeotropic anchoring, the director field must respond to complex geometric constraints. In this system, experimental and theoretical research has found rich structural behaviors, which depend on the Frank elastic constants, surface anchoring strength, radius of the cylinder, defect core size, and external fields~\cite{allender1991determination,kralj1993stability,kralj1995saddle,burylov1997equilibrium,shams2014theoretical}. When the surface anchoring is strong, external fields are absent, the the cylinder radius is large enough compared to the defect core, the equilibrium configuration is a defect-free director field, where the director gradually escapes into the third dimension, moving from the boundary to the center of the cylinder. As the cylinder radius decreases, at a certain point, a configuration with two $+1/2$ wedge disclination lines parallel to the cylindrical axis becomes more stable than the escaped radial configuration~\cite{de2007point}. The positions of those two $+1/2$ disclination lines are affected by both the interaction between the two disclinations and the repulsion from the boundaries. In this section, we apply the Peach-Koehler force theory to understand the equilibrium positions of those two disclination lines inside the cylinder.

\begin{figure}
\centering
\includegraphics[width=.7\textwidth]{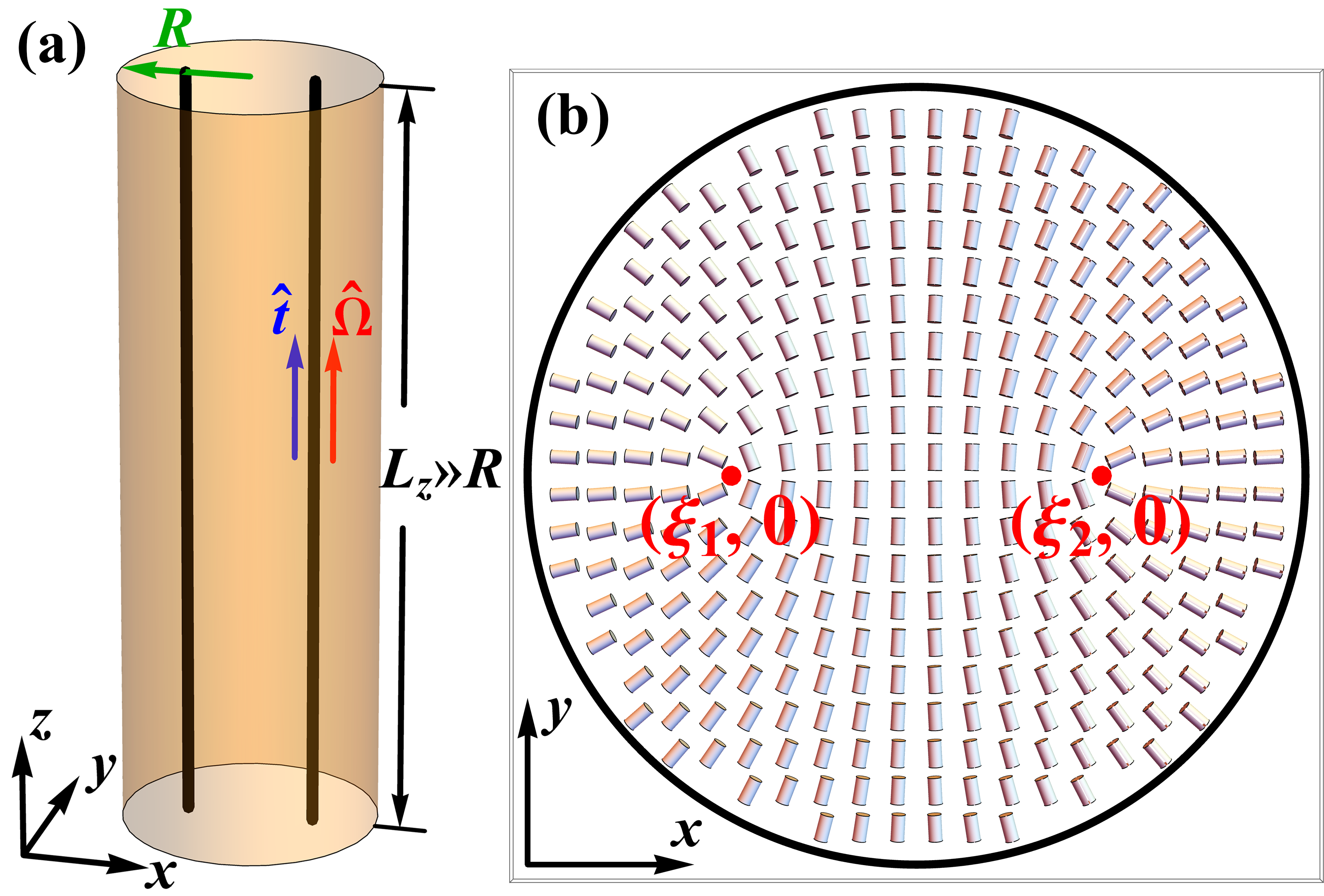}
\caption{(a)~Schematic illustration of a nematic liquid crystal inside a cylinder with homeotropic anchoring, with two $+1/2$ wedge disclination lines running parallel to the cylindrical axis. (b)~Director field in the $x$-$y$ plane, with disclinations located at $(\xi_1,0)$ and $(\xi_2,0)$.} \label{fig5:twoplushalfdisclinations}
\end{figure}

We consider a nematic liquid crystal confined in a long cylinder of radius $R$, with strong homeotropic anchoring on the side of the cylinder, as shown in Figure~\ref{fig5:twoplushalfdisclinations}(a). We assume that the stable state is two $+1/2$ wedge disclination lines in the cylinder, based on the ratio of the cylinder radius to the defect core size. The axial direction of the cylinder is aligned with the $z$-axis, and the length of the cylinder is $L_z$. We assume $L_z \gg R$, and the system is uniform along the $z$-axis. The director field inside the cylinder is thus independent of $z$, denoted by $\hat{\boldsymbol{n}}(x,y)$, and one can simplify the 3D cylindrical model into a 2D disk with homeotropic anchoring in the $x$-$y$ plane, shown in Figure~\ref{fig5:twoplushalfdisclinations}(b). Because of the surface anchoring, we further assume the director inside the cylinder is confined in the $x$-$y$ plane, so that $\hat{\boldsymbol{n}} = \left( \cos{\phi}, \sin{\phi}, 0 \right)$, as in Eq.~(\ref{sec2:director}). The free energy is still given by Eq.~(\ref{sec2:freeenergy}), and the Euler-Lagrange equation by Eq.~(\ref{sec2:pdeforphi}).

We assume that the two $+1/2$ disclinations are on the same diameter across the cylinder, which we identify as the $x$-axis, and we label their positions as $x=\xi_1$ and $x=\xi_2$. In that case, the director field has a mirror symmetry about the $x$-$z$ plane. Using that symmetry, we can reduce the problem of calculating the director field from the full disk to the upper half of the disk, leaving the singularity points on the boundaries. On this upper half disk, the boundary conditions can be written as $\phi(x,0)=0$ for $\xi_2<x<R$, $\phi(x,0)=\pi/2$ for $\xi_1<x<\xi_2$, $\phi(x,0)=\pi$ for $-R<x<\xi_1$, and $\phi(x,y)=\arctan{(y/x)}$ for $x^2+y^2=R^2$.

We solve the Euler-Lagrange equation for $\phi(x,y)$ with those boundary conditions by conformal mapping, and obtain
\begin{align}\label{sec4:phisolution}
  \phi(x,y) &= \frac{1}{2} \arctan{\left( -x + \frac{R^2}{\xi_1} + \xi_1 - \frac{x R^2}{x^2+y^2}, -y + \frac{y R^2}{x^2+y^2} \right)}\\
  &\quad + \frac{1}{2} \arctan{\left( -x + \frac{R^2}{\xi_2} + \xi_2 - \frac{x R^2}{x^2+y^2}, -y + \frac{y R^2}{x^2+y^2} \right)}
  + \arctan{\left( x,y \right)} -\frac{\pi}{2}.\nonumber
\end{align}
Here, $\arctan{(x,y)}$ is defined as the angle between vectors $(x,y)$ and $(1,0)$. When $(x,y)$ is in the upper half $x$-$y$ plane, $\arctan{(x,y)}$ has positive values; when $(x,y)$ is in the lower half plane, it has negative values; the return value is in the range from $-\pi$ to $\pi$. Those solutions all satisfy the Euler-Lagrange equation, as well as the boundary conditions on the entire disk. As in the previous sections, this solution applies for all possible positions of the disclinations, $\xi_1$ and $\xi_2$. Here, we will determine the the equilibrium values of $\xi_1$ and $\xi_2$ by calculating the Peach-Koehler forces on the disclinations.

First, let us determine the Peach-Koehler force on disclination \#2. We combine the analytic solution of Eq.~(\ref{sec4:phisolution}) with the previous Eqs.~(\ref{sec2:director}) and~(\ref{sec2:sigmatotal}) to calculate the total angular stress tensor $\boldsymbol{\sigma}^{\mathrm{total}}$. We change the coordinates from $(x,y)$ to polar coordinates $(\rho,\varphi)$ centered on the disclination, 
\begin{equation}\label{sec4:coordtransformation}
    x = \xi_2 + \rho \cos{\varphi}, \quad
    y = \rho \sin{\varphi}.
\end{equation}
We expand the total angular stress as a power series around $\rho=0$ to obtain
\begin{equation}\label{sec4:totalstress}
  \boldsymbol{\sigma}^{\mathrm{total}} = K
  \begin{pmatrix}
  0 & 0 & 0 \\
  0 & 0 & 0 \\
  -\dfrac{\sin{\varphi}}{2\rho} & \dfrac{\cos{\varphi}}{2\rho} + \dfrac{1}{2}\left(\dfrac{1}{\xi_2 - \xi_1} + \dfrac{\xi_2}{\xi_2^2 - R^2} + \dfrac{\xi_1}{\xi_1\xi_2 - R^2} \right) & 0
  \end{pmatrix} + O(\rho).
\end{equation}
From this total angular stress, we must subtract the angular stress of a free $+1/2$ wedge disclination located at $(\xi_2,0)$. This free disclination has the director field
\begin{equation}
  \phi_{\mathrm{free}}(x,y) = \frac{1}{2}\arctan{(x-\xi_2,y)},
\end{equation}
and the angular stress tensor
\begin{equation}
  \boldsymbol{\sigma}^{\mathrm{free}} = K
  \begin{pmatrix}
    0 & 0 & 0 \\
    0 & 0 & 0 \\
    -\dfrac{\sin{\varphi}}{2\rho} & \dfrac{\cos{\varphi}}{2\rho} & 0
  \end{pmatrix}.
\end{equation}
Hence, the external angular stress acting on the disclination at $\xi_2$ is
\begin{equation}
  \boldsymbol{\sigma}^{\mathrm{ext}} = \lim_{\rho \to 0} \left( \boldsymbol{\sigma}^{\mathrm{total}} - \boldsymbol{\sigma}^{\mathrm{free}} \right) = K
  \begin{pmatrix}
    0 & 0 & 0 \\
    0 & 0 & 0 \\
    0 & \dfrac{1}{2}\left( \dfrac{1}{\xi_2 - \xi_1} + \dfrac{\xi_2}{\xi_2^2 - R^2} + \dfrac{\xi_1}{\xi_1\xi_2 - R^2} \right) & 0
  \end{pmatrix}.
\end{equation}
For the Peach-Koehler force, we must combine $\boldsymbol{\sigma}^{\mathrm{ext}}$ with the tangent vector $\hat{\boldsymbol{t}}$ and rotation vector $\hat{\boldsymbol{\Omega}}$ in Eq.~(\ref{sec1:FPK}). Because the disclination line is aligned with the $z$-axis, we can choose $\hat{\boldsymbol{t}}=(0,0,1)$. For a $+1/2$ wedge disclination, we also have $\hat{\boldsymbol{\Omega}}=(0,0,1)$. Hence, the Peach-Koehler force acting on disclination \#2 becomes
\begin{equation}\label{sec4:pkforceonxi2}
  \boldsymbol{F}_{\mathrm{PK}}^{(2)} = \left( \frac{\pi K}{2} \left( \frac{1}{\xi_2 - \xi_1} + \frac{1}{\xi_2 - R^2/\xi_2} + \frac{1}{\xi_2 - R^2/\xi_1} \right), 0, 0 \right).
\end{equation}

To interpret this force, we note that it has no $z$ component because of mirror symmetry about the $x$-$y$ plane, and no $y$ component because of mirror symmetry about the $x$-$z$ plane. In the $x$ component, the first term shows the force on disclination \#2 arising from its repulsive interaction with disclination \#1. It scales as $K/(\xi_2 - \xi_1)$, consistent with the usual repulsive force between two disclinations of topological charge $+1/2$~\cite{long2021geometry}. The remaining two terms represent the interaction between disclination \#2 and the boundary constraints. They can be understood as interactions with image defects located at $R^2/\xi_1$ and $R^2/\xi_2$, as in electrostatic problems.

The Peach-Koehler force acting on the other disclination \#1 can be found by exactly the same argument, swapping the labels 1 and 2, which gives
\begin{equation}\label{sec4:pkforceonxi1}
  \boldsymbol{F}_{\mathrm{PK}}^{(1)} = \left( \frac{\pi K}{2} \left( \frac{1}{\xi_1 - \xi_2} + \frac{1}{\xi_1 - R^2/\xi_1} + \frac{1}{\xi_1 - R^2/\xi_2} \right), 0, 0 \right).
\end{equation}
Its interpretation is analogous to that for disclination \#2.

To find the equilibrium positions for the two disclinations, we solve simultaneously the two equations $\boldsymbol{F}_{\mathrm{PK}}^{(2)}=0$ and $\boldsymbol{F}_{\mathrm{PK}}^{(1)}=0$. The result is
\begin{equation}
    \xi_1 = - \frac{R}{5^{1/4}}, \quad
    \xi_2 = \frac{R}{5^{1/4}}.
\end{equation}
Vafa et al.\ recently investigated a similar but more general problem for topological defects on a cone~\cite{vafa2022defect}. In the limiting case when the cone is a flat disk, their prediction for the optimal distance between two $+1/2$ defects agrees with our findings here.

\section{Radial hedgehog or disclination loop in a spherical nematic droplet with homeotropic anchoring}

\subsection{No magnetic field}

Hedgehogs are point defects in 3D nematic liquid crystals~\cite{kleman2006topological}. In a radial hedgehog, for example, the director points radially outward away from the defect in all directions. Far from a radial hedgehog, the director field is actually equivalent to the director field around a $+1/2$ wedge disclination loop. Hence, a point hedgehog can expand into a disclination loop, or conversely, a disclination loop can contract into a point hedgehog. Because of this relationship, one must ask: What is the equilibrium structure---a point hedgehog or a disclination loop? If it is a loop, what is the equilibrium radius?

There have been several theoretical and experimental studies of this issue~\cite{kralj1992freedericksz,terentjev1995disclination,fukuda2002stability,wang2016experimental}. Using oblate spheroidal coordinates, Terentjev found that in infinite space, the equilibrium structure of a radial hedgehog is a $+1/2$ wedge disclination loop with a stable radius of $r_d \sim 30 r_c$, where $r_c$ is the disclination core radius~\cite{terentjev1995disclination}. Likewise, Fukuda and Yokoyama calculated that the equilibrium structure of a hyperbolic hedgehog is a $-1/2$ wedge disclination loop with $r_d \sim 3 r_c$~\cite{fukuda2002stability}. Later, using experiments that couple liquid crystal topology with amphiphilic self-assembly, Wang et al.\ observed that the equilibrium structure of a radial hedgehog is a disclination loop with radius of about $40$ nm~\cite{wang2016experimental}. In this section, we investigate the equilibrium structure of a radial hedgehog or disclination loop using the Peach-Koehler force theory.

\begin{figure}
\centering
\includegraphics[width=.5\textwidth]{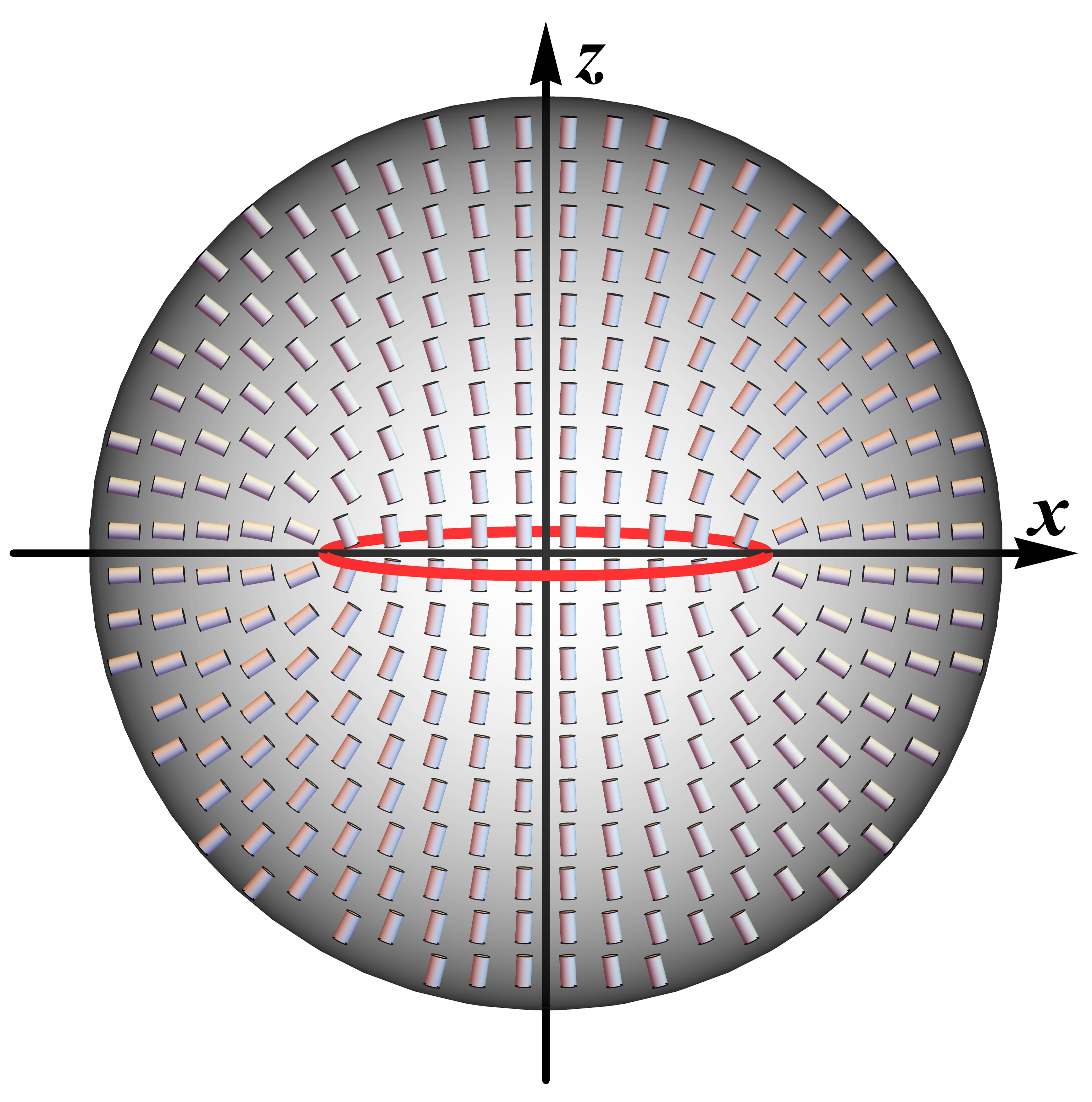}
\caption{Nematic liquid crystal inside a spherical droplet of radius $R$, with homeotropic anchoring. The red line represents a $+1/2$ wedge disclination loop of radius $r_d$. The director field has continuous rotational symmetry around the $z$-axis. (For clarity, the director field is only shown in the $x$-$z$ plane.)} \label{fig6}
\end{figure}

We consider a liquid crystal inside a spherical droplet of radius $R$, with strong homeotropic anchoring, as shown in Figure~\ref{fig6}. In this geometry, the director field must have a radial hedgehog or $+1/2$ wedge disclination loop inside the droplet. We suppose that there is a $+1/2$ wedge disclination loop of radius $r_d$, and investigate whether the equilibrium structure has $r_d\to0$ or $r_d$ finite.

For this calculation, we suppose that the disclination loop is in the $x$-$y$ plane, and shares the same center with the spherical droplet. Because the director field has azimuthal symmetry around the $z$-axis, it is easiest to describe the director field in cylindrical coordinates $(r,\theta,z)$. These coordinates define a set of orthonormal basis vectors, $\hat{\boldsymbol{r}}$, $\hat{\boldsymbol{\theta}}$, and $\hat{\boldsymbol{z}}$. Based on symmetry, it is plausible to assume the director field is everywhere in the $\hat{\boldsymbol{r}}$-$\hat{\boldsymbol{z}}$ plane. Hence, it can thus be described by a single angle $\phi(r, z)$ as
\begin{equation}\label{sec5:directorbyphi}
  \hat{\boldsymbol{n}}\left( r, z \right) = \hat{\boldsymbol{r}} \cos{\phi} + \hat{\boldsymbol{z}} \sin{\phi}.
\end{equation}
For a nematic liquid crystal with a single Frank elastic constant $K$, the total free energy associated with this director field $\hat{\boldsymbol{n}}(r,z)$ is given by
\begin{equation}\label{sec6:totalfreeenergy}
  F = \int 2 \pi r\, \mathrm{d}r\, \mathrm{d}z \; \frac{1}{2}K\left[ \left( \partial_{r}\phi \right)^2 + \left( \partial_{z}\phi \right)^2 + \frac{\cos^2{\phi}}{r^2} \right].
\end{equation}
The corresponding Euler-Lagrange equation for $\phi(r,z)$ becomes
\begin{equation}
  -2 \pi K \left( r \partial_{r}^2 \phi + r \partial_{z}^2 \phi + \partial_{r}\phi + \frac{\sin{2\phi}}{2r} \right) = 0.
\end{equation}

In principle, if we want to analyze the exact Peach-Koehler force acting on the disclination loop, we need to solve this partial differential equation analytically with the homeotropic boundary condition. However, unlike the cases in the previous sections, the Euler-Lagrange equation for the angle $\phi(r,z)$ is a nonlinear partial differential equation which cannot be solved exactly. Instead, we make an approximation: We just use the solution of $\phi(x,y)$ from Eqn.~(\ref{sec4:phisolution}) in Section 4, substituting coordinates $(r,z)$ for $(x,y)$, and assuming $\xi_1 = -\xi_2 = r_d$. This approximation satisfies the homeotropic boundary condition exactly, although it does not satisfy the Euler-Lagrange equation exactly. We will use this approximation for the rest of the calculation.

To calculate the Peach-Koehler force, let us consider the point where the disclination loop crosses the $x$-axis, which is $(r_d,0,0)$. As in the previous sections, we need to inspect the angular stress tensor near the disclination, in the plane that is locally perpendicular to the disclination, which is the $x$-$z$ plane. In this local plane, we can use local polar coordinates $(\rho, \varphi)$ centered on the disclination, defined by
\begin{equation}
    x = r_d + \rho \cos{\varphi}, \quad
    y = 0, \quad
    z = \rho \sin{\varphi}.
\end{equation}
The power series of $\boldsymbol{\sigma}^{\mathrm{total}}$ expanded around $\rho=0$ is
\begin{equation}
  \boldsymbol{\sigma}^{\mathrm{total}} = K
    \begin{pmatrix}
      0 & - \dfrac{\sin{\varphi}}{2 r_d} &  0 \\
      \dfrac{\sin{\varphi}}{2 \rho} & 0 & - \dfrac{\cos{\varphi}}{2 \rho}
      - \dfrac{R^4 -5 r_d^4}{4 r_d (R^4 - r_d^4)} \\
      0 & \dfrac{1+\cos{\varphi}}{2r_d} & 0
    \end{pmatrix}
    + O(\rho).
\end{equation}
The divergent term in $\rho$ is the angular stress induced by a free $+1/2$ wedge disclination, and it is not related to the disclination loop radius $r_d$ or the nematic droplet radius $R$. We subtract it from the total angular stress, and then take the limit $\rho\to0$, to obtain the external angular stress acting on the disclination
\begin{equation}
  \boldsymbol{\sigma}^{\mathrm{ext}} = \lim_{\rho \to 0} \left( \boldsymbol{\sigma}^{\mathrm{total}} - \boldsymbol{\sigma}^{\mathrm{free}} \right) = K
    \begin{pmatrix}
      0 & - \dfrac{\sin{\varphi}}{2 r_d} &  0 \\
      0 & 0 & - \dfrac{R^4 -5 r_d^4}{4 r_d (R^4 -r_d^4)} \\
      0 & \dfrac{1+\cos{\varphi}}{2r_d} & 0
    \end{pmatrix}.
\end{equation}
We must now combine $\boldsymbol{\sigma}^{\mathrm{ext}}$ with the tangent vector $\hat{\boldsymbol{t}}$ and the rotation vector $\hat{\boldsymbol{\Omega}}$ in Eq.~(\ref{sec1:FPK}) to calculate the Peach-Koehler force. At the point under consideration, where the disclination loop crosses the $x$-axis, the tangent vector can be chosen as $\hat{\boldsymbol{t}}=(0, 1, 0)$. For a $+1/2$ wedge disclination, the corresponding rotation vector is also $\hat{\boldsymbol{\Omega}}=(0, 1, 0)$. The resulting Peach-Koehler force is in the $\hat{\boldsymbol{x}}$ direction, or equivalently in the $\hat{\boldsymbol{r}}$ direction, radially outward from the $z$ axis. By symmetry, the Peach-Koehler force must be radially outward \emph{everywhere} around the disclination loop, and hence it can be written as
\begin{equation}\label{sec5:pkforce}
  \boldsymbol{F}_{\mathrm{PK}} = \hat{\boldsymbol{r}} \frac{\pi K \left(R^4 - 5 r_d^4 \right)}{4 r_d \left(R^4 - r_d^4 \right)}.
\end{equation}

To interpret this result, we note that the Peach-Koehler force comes from two sources: the self-interaction of the disclination loop because of its curved shape, and the interaction between the disclination loop and the spherical boundary. When $r_d\ll R$, the disclination loop becomes so small that the self-interaction dominates over the repulsive force from the boundary. The Peach-Koehler force is then equivalent to the repulsive force between two infinite straight $+1/2$ disclination lines,
\begin{equation}\label{sec5:selfinteraction}
  \boldsymbol{F}_{\mathrm{PK}} \approx \hat{\boldsymbol{r}} \frac{\pi K}{4 r_d}.
\end{equation}
By comparison, when $r_d$ approaches $R$, the Peach-Koehler force is dominated by the strong repulsive force from the wall, and it becomes
\begin{equation}
  \boldsymbol{F}_{\mathrm{PK}} \approx - \hat{\boldsymbol{r}} \frac{\pi K}{4 (R-r_d)}.
\end{equation}

Unlike the examples in previous sections, the Peach-Koehler force is not enough to determine the equilibrium size of the disclination loop. In addition to the Peach-Koehler force, we must also consider the line tension of the disclination. The disclination core is a region where the bulk uniaxial nematic order is disrupted, and the nematic order tensor must have different eigenvalues. Suppose that $I$ is the free energy cost of the disclination core per length. The total core free energy of the loop is then $2\pi I r_d$, and the force per length acting on the disclination is
\begin{equation}
  \boldsymbol{F}_{\mathrm{tension}} = - \hat{\boldsymbol{r}} \frac{I}{r_d}.
\end{equation}
This tension force compresses the disclination loop to be smaller, and the magnitude of this force increases as $r_d$ decreases. By comparing the tension force with the self-interaction of the disclination loop in Eq.~(\ref{sec5:selfinteraction}), we can see that if $I/K>\pi/4$, the tension force would drag the disclination loop down to a point. That result seems to contradict previous findings that a free radial hedgehog is usually a $+1/2$ wedge disclination loop with small radius~\cite{terentjev1995disclination,wang2016experimental}.

To resolve this contradiction, we suggest that one more force is needed. The $+1/2$ wedge disclination loop does not have a uniform rotation vector $\hat{\boldsymbol{\Omega}}$. Rather, $\hat{\boldsymbol{\Omega}}$ points tangentially around the loop, and hence it varies as a function of position. This variation of $\hat{\boldsymbol{\Omega}}$ must cost free energy, which can be regarded as a bending energy for the disclination line. If we were to do a Landau-de Gennes theory of the disclination loop, the bulk nematic phase would be a positive uniaxial region with order parameter $s_0$ and director $\hat{\boldsymbol{n}}$, while the defect core would be a negative uniaxial region with order parameter $-s_0/2$ and director $\hat{\boldsymbol{\Omega}}$~\cite{long2021geometry}. Hence, we estimate that the elastic constant for variation of $\hat{\boldsymbol{\Omega}}$ should be $K/4$, and the associated free energy density of the disclination core should be $K/(8r_d)^2$. The core has volume $(2\pi r_d)(\pi r_c^2)$, where $r_c$ is the core radius. Hence, we estimate that the total free energy associated with variation of $\hat{\boldsymbol{\Omega}}$ should be
\begin{equation}\label{sec5:freeenergyforomega}
  F_{\Omega} = \frac{\pi^2 K r_c^2}{4 r_d}.
\end{equation}
The corresponding force per length acting on the disclination loop due to the spatially varying $\hat{\boldsymbol{\Omega}}$ is
\begin{equation}
  \boldsymbol{F}_{\Omega} = \hat{\boldsymbol{r}} \frac{\pi K r_c^2}{8 r_d^3}.
\end{equation}
This force pushes radially outward on the disclination loop. Because it scales as $1/r_d^3$, it prevents the loop from shrinking to a point due to the tension force, even if $I/K > \pi/4$. We conjecture that this force should also occur in other geometries, such as hyperbolic hedgehogs.

\begin{figure}
\centering
\includegraphics[width=1\textwidth]{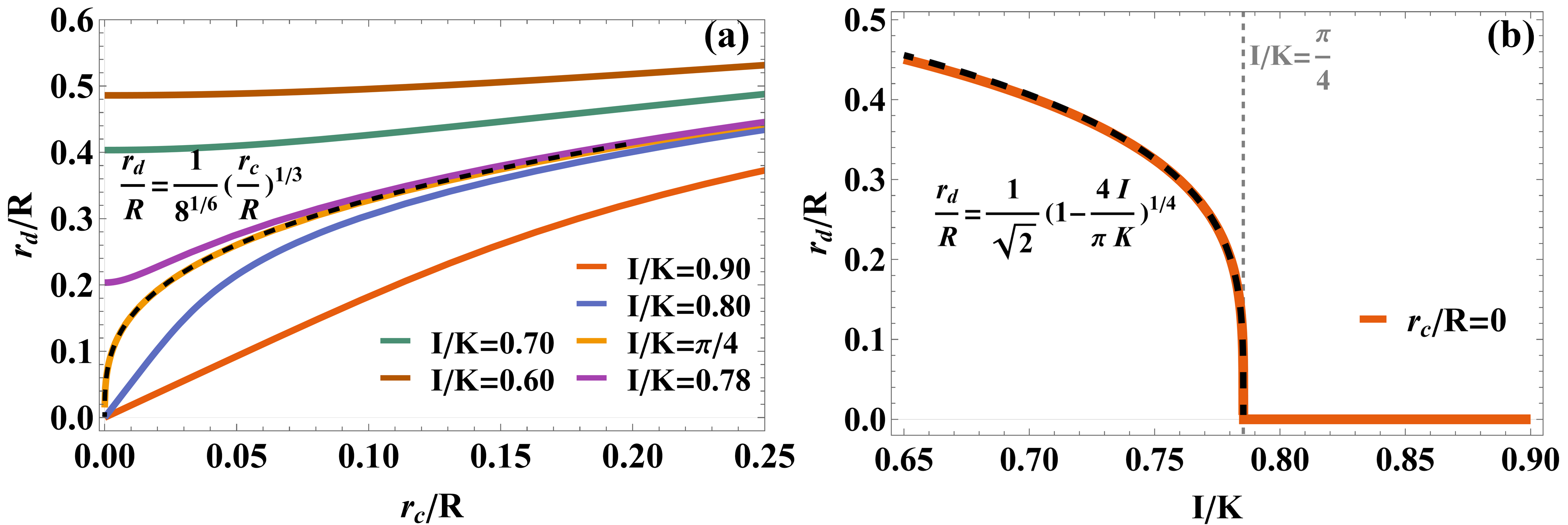}
\caption{Predictions for a $+1/2$ wedge disclination loop in a spherical nematic droplet, with no magnetic field. (a)~Equilibrium loop radius $r_d$ as a function of defect core radius $r_c$, at different ratios of disclination line tension $I$ to elastic constant $K$, solved numerically from $\boldsymbol{F}_{\mathrm{total}} = 0$. When $I/K=\pi/4$, the plot follows the critical form of Eq.~(\ref{sec5:loopradiuscritical}), shown by the black dashed line. When $I/K<\pi/4$, the plot shifts upward to a nonzero value even in the limit $r_c\to0$. (b)~Equilibrium loop radius $r_d$ as a function of $I/K$ at $r_c/R=0$. The orange solid line shows the numerical solution, and the black dashed line shows the asymptotic form of Eq.~(\ref{sec5:loopradiusasymptotic}).} \label{fig7:numsolutionforrd}
\end{figure}

Combining all the forces including $\boldsymbol{F}_{\mathrm{PK}}$, $\boldsymbol{F}_{\mathrm{tension}}$, and $\boldsymbol{F}_{\Omega}$, the total force acting on the disclination loop is given by
\begin{equation}
  \boldsymbol{F}_{\mathrm{total}} = \hat{\boldsymbol{r}} \frac{K}{R} \left( \frac{\pi\left( 1-5\tilde{r}_d^4 \right)}{4 \tilde{r}_d \left( 1-\tilde{r}_d^4 \right)} - \frac{\tilde{I}}{\tilde{r}_d} + \frac{\pi \tilde{r}_c^2}{8\tilde{r}_d^3} \right),
\end{equation}
where $\tilde{r}_d \equiv r_d/R$, $\tilde{r}_c \equiv r_c/R$, and $\tilde{I} \equiv I/K$. The equilibrium radius of the disclination loop can be determined from $\boldsymbol{F}_{\mathrm{total}}=0$. We can see that the dimensionless equilibrium radius $\tilde{r}_d$ depends only on the dimensionless core radius $\tilde{r}_c$, and on the dimensionless ratio $\tilde{I}$ of line tension to elastic constant. Figure~\ref{fig7:numsolutionforrd}(a) shows numerical results for the relation between $\tilde{r}_d$ and $\tilde{r}_c$ at different values of $\tilde{I}$. In these results, we can distinguish three cases:

\emph{Case 1.} For $\tilde{I} > \pi/4$, the tension force is strong enough to overcome the repulsive self-interaction of the defect loop. The equilibrium loop radius then approaches $\tilde{r}_d = 0$ as $\tilde{r}_c$ decreases to zero, and expands monotonically as $\tilde{r}_c$ increases. Applying perturbation theory for $\tilde{I} > \pi/4$ and $\tilde{r}_c \ll 1$, the loop size is approximately
\begin{equation}\label{sec5:loopradiusperturbation}
  r_d = \frac{r_c}{\sqrt{2\left( \frac{4 I}{\pi K} - 1 \right)}}.
\end{equation}
This result shows that the equilibrium loop radius is proportional to the disclination core radius, and the ratio is determined by $\tilde{I}$. In this approximation, the equilibrium loop radius is independent of the droplet size $R$. This independence is understandable because the size of the disclination loop is mainly determined by the balance between  $1/r_d$ forces (from $\boldsymbol{F}_{\mathrm{tension}}$ and from self-repulsion in $\boldsymbol{F}_{\mathrm{PK}}$) and $1/r_d^3$ force (from $\boldsymbol{F}_{\Omega}$). It is not related to force from the boundary in $\boldsymbol{F}_{\mathrm{PK}}$. We conjecture that this case is the most common in experiments, where a radial hedgehog is often observed as a very small ring, with radius comparable to $r_c$, or even as a point defect.

\emph{Case 2.} For $\tilde{I}=\pi/4$, the derivative of $r_d$ with respect to $r_c$ becomes divergent at the origin.  The critical behavior of $r_d$ can then be calculated as
\begin{equation}\label{sec5:loopradiuscritical}
  \frac{r_d}{R} = \frac{1}{8^{1/6}} \left( \frac{r_c}{R} \right)^{1/3}.
\end{equation}
In this case, the the disclination loop radius becomes dependent on the nematic droplet radius $R$. This dependence is reasonable, because $\boldsymbol{F}_{\mathrm{tension}}$ exactly cancels the self-repulsion in $\boldsymbol{F}_{\mathrm{PK}}$, and hence the loop radius is determined by the balance between the outward repulsion of $\boldsymbol{F}_{\Omega}$ and the inward force from the boundary in $\boldsymbol{F}_{\mathrm{PK}}$.

\emph{Case 3.} For $\tilde{I}<\pi/4$, the inward tension force is not strong enough to cancel the outward force from the self-repulsion of the disclination loop. Hence, the equilibrium size of the disclination loop remains finite even if $r_c = 0$. Figure~\ref{fig7:numsolutionforrd}(b) shows the numerical solution for the dimensionless loop radius $\tilde{r}_d$ as a function of $\tilde{I}$ at $\tilde{r}_c=0$. For $\tilde{I}$ slightly smaller than $\pi/4$, the asymptotic behavior of $\tilde{r}_d$ as a function of $\tilde{I}$ is
\begin{equation}\label{sec5:loopradiusasymptotic}
  \frac{r_d}{R} = \frac{1}{\sqrt{2}} \left( 1-\frac{4 I}{\pi K} \right)^{1/4}.
\end{equation}
In this limit of $r_c = 0$, the equilibrium loop radius $r_d$ is linearly proportional to the droplet radius $R$, because there is no other length scale.

\subsection{Magnetic field}

Ettinger et al.~\cite{ettinger2023magnetic} have recently performed experiments on $+1/2$ wedge disclination loops in a spherical nematic droplets under applied magnetic fields. Inspired by these experiments, we now repeat the Peach-Koehler force theory above, but with an applied magnetic field.

From previous experimental and computational studies~\cite{kralj1992freedericksz,wang2016experimental,ettinger2023magnetic}, the director field of a nematic droplet under a magnetic field has the same general form shown in Figure~\ref{fig6}, and we again represent it by Eq.~(\ref{sec5:directorbyphi}) in terms of the single angle $\phi(r,z)$. The applied field $\boldsymbol{B}=B\hat{\boldsymbol{z}}$ now defines the symmetry axis $\hat{\boldsymbol{z}}$. We consider a liquid crystal with positive diamagnetic anisotropy, so that the director tends to align parallel to the field. This aligning effect can be represented by an extra term in the free energy density of $-(\chi_a/(2\mu_0))(\hat{\boldsymbol{n}}\cdot \boldsymbol{B})^2$, where $\chi_a$ is the diamagnetic anisotropy. To characterize the effect of the applied magnetic field on the director field, it is convenient to define a magnetic coherence length $\xi_{\mathrm{mag}} = B^{-1}\sqrt{K\mu_0/\chi_a}$. The total free energy associated with the magnetic field then becomes
\begin{equation}\label{sec5:magneticfieldenergy}
  F_{H} = \int 2 \pi r\,\mathrm{d}r\, \mathrm{d}z \left( - \frac{1}{2} \frac{K}{\xi_{\mathrm{mag}}^2} \sin^2{\phi}\right),
\end{equation}
which generates one more nonlinear term in the Euler-Lagrange equation for $\phi(r,z)$.

In principle, we must solve the new Euler-Lagrange equation for $\phi(r,z)$ to calculate the forces acting on the disclination loop. Here, as a much simpler alternative, we take the solution from Eq.~(\ref{sec4:phisolution}) again as an approximation for the director field inside the nematic droplet, and hence we obtain the Peach-Koehler force directly from our previous calculation. We also use $\boldsymbol{F}_{\mathrm{tension}}$ and $\boldsymbol{F}_{\Omega}$ found above. We still need to determine the extra force imposed by the magnetic field.

By symmetry, the magnetic force on the disclination loop can only be outward or inward, parallel or antiparallel to $\hat{\boldsymbol{r}}$ in the cylindrical coordinates. From Figure~\ref{fig6}, we can see that the director field inside the loop is almost uniformly aligned with the magnetic field $\boldsymbol{B}=B\hat{\boldsymbol{z}}$, while the director field outside the disclination loop is highly nonuniform. Hence, the region inside the loop has a lower magnetic free energy that the region outside. For that reason, the magnetic force must push outward, parallel to $\hat{\boldsymbol{r}}$, to expand the loop. To estimate the strength of the magnetic force, we use dimensional analysis. Like the Peach-Koehler force, the magnetic force is really a force per length acting on the disclination. Because it is derived from a change in the magnetic free energy of Eqn.~(\ref{sec5:magneticfieldenergy}) as the disclination moves, it must depend on the parameters $K$ and $\xi_{\mathrm{mag}}$.  The combination $K/\xi_{\mathrm{mag}}$ has the appropriate dimensions. Hence, as a rough estimate, we will just write the magnetic force as
\begin{equation}
  \boldsymbol{F}_{\mathrm{mag}} = \hat{\boldsymbol{r}} \frac{K}{\xi_{\mathrm{mag}}} = \hat{\boldsymbol{r}} \frac{K}{R \tilde{\xi}_{\mathrm{mag}}},
\end{equation}
where $\tilde{\xi}_{\mathrm{mag}} \equiv \xi_{\mathrm{mag}}/R$ is defined as a dimensionless magnetic coherence length. This magnetic force is zero when the applied field is zero, and increases as the field increases.

\begin{figure}
\centering
\includegraphics[width=1\textwidth]{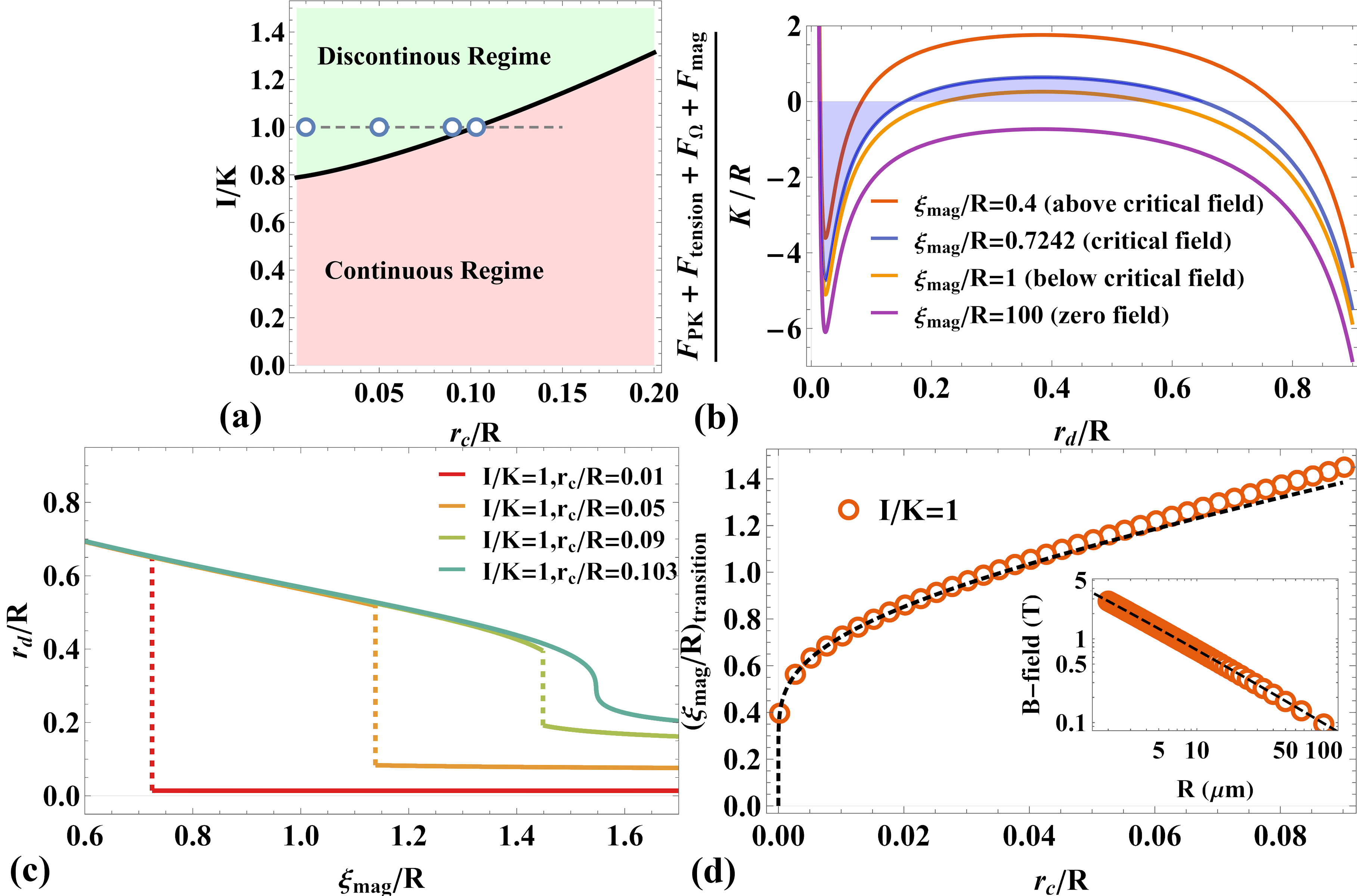}
\caption{Predictions for a $+1/2$ wedge disclination loop in a spherical nematic droplet, under an applied magnetic field. (a)~Parameter space of $I/K$ and $r_c/R$, showing whether the equilibrium loop radius increases continuously or discontinuously as a function of magnetic field. The four blue circles corresponds to the four sets of parameters in section~(c). The gray dashed line shows the parameters in section~(d). (b)~Total force per length acting on the loop as a function of $r_d/R$, at different magnetic field strengths, when $I/K=1$ and $r_c/R=0.01$. The blue region (between the blue force curve and 0) shows the integral of total force between the two stable points where total force vanishes. This integral is zero at the transition field. (c)~Equilibrium loop radius $r_d/R$ as a function of $\xi_{\mathrm{mag}}/R$ (related to magnetic field) at different values of $r_c/R$, for $I/K=1$. (d)~Value of $\xi_{\mathrm{mag}}/R$ at the first-order transition, as a function of $r_c/R$, at $I/K=1$. The black dashed line indicates the logarithmic behavior of $r_c/R$ close to zero. The inset figure shows the corresponding magnetic field in Tesla at the transition, as a function of droplet size $R$ in microns, assuming $r_c = 10$~nm, $I = K = 10^{-11}$~N, and $\chi_a = 10^{-6}$.} \label{fig8:firstordertransition}
\end{figure}

We now combine all the forces, $\boldsymbol{F}_{\mathrm{PK}}$, $\boldsymbol{F}_{\mathrm{tension}}$, $\boldsymbol{F}_{\mathrm{\Omega}}$, and $\boldsymbol{F}_{\mathrm{mag}}$, to determine the equilibrium radius of a disclination loop as a function of applied magnetic field. We explore the parameter space of $\tilde{I}$ and $\tilde{r}_c$, and find two qualitatively different behaviors for the response to magnetic field, shown in Figure~\ref{fig8:firstordertransition}(a). When $\tilde{I}$ and $\tilde{r}_c$ are in the continuous regime, the loop radius increases continuously as the magnetic field increases. In the limit of very strong field, the disclination loop is eventually pushed to the spherical boundary where $\tilde{r}_d \approx 1$. By comparison, when $\tilde{I}$ and $\tilde{r}_c$ are in the discontinuous regime, the loop radius exhibits a discontinuous jump at a transition field strength, which depends on $\tilde{I}$ and $\tilde{r}_c$.

To understand why the discontinuous transition occurs, consider the example parameters $\tilde{I} = 1$ and $\tilde{r}_c = 0.01$. Figure~\ref{fig8:firstordertransition}(b) shows plots of the total force as a function of loop radius $r_d$, for several values of $\tilde{\xi}_{\mathrm{mag}}$ (i.e.\ several values of magnetic field). For large $\tilde{\xi}_{\mathrm{mag}} = 100$ (magnetic field almost zero), the total force curve crosses zero at only one point, which is the small loop radius $\tilde{r}_d \approx 0.014$. As $\tilde{\xi}_{\mathrm{mag}}$ decreases (magnetic field increases), the magnetic force causes the total force curve to shift upward. Eventually, the total force curve crosses zero at three points. The first and third points are stable, and the second is unstable. To determine which of the stable points is the absolute minimum of free energy, we calculate the free energy difference by integrating the total force curve between those points. For $\tilde{\xi}_{\mathrm{mag}} = 1$, the leftmost point has the lowest free energy, so the equilibrium loop radius is still small. For $\tilde{\xi}_{\mathrm{mag}} = 0.7242$, the free energy difference between those two stable points becomes zero. Here, the system is exactly at the transition magnetic field. For smaller $\tilde{\xi}_{\mathrm{mag}}$ (larger magnetic field), the rightmost point has the lowest free energy, so the equilibrium loop radius is large, comparable to the droplet size $R$. Eventually, when $\tilde{\xi}_{\mathrm{mag}}$ is sufficiently small (magnetic field sufficiently large), the total force curve only crosses zero at one point, which is a large loop radius. Thus, this example explicitly shows a discontinuous, first-order transition from small to large loop radius at a transition field.

To explore how the discontinuous transition evolves into continuous behavior, we plot the equilibrium loop radius $\tilde{r}_d$ as a function of $\tilde{\xi}_{\mathrm{mag}}$. Figure~\ref{fig8:firstordertransition}(c) shows several plots at different points in the $\tilde{I}$-$\tilde{r}_c$ parameter space. The red plot with $\tilde{I} = 1$ and $\tilde{r}_c = 0.01$ corresponds to the behavior discussed in the previous paragraph. In this example, as $\tilde{\xi}_{\mathrm{mag}}$ decreases through $0.7242$, the equilibrium loop radius jumps discontinuously from 
from $\tilde{r}_d \approx 0.014$ to $\tilde{r}_d \approx 0.65$. As $\tilde{r}_c$ increases, the repulsive force $\boldsymbol{F}_{\mathrm{\Omega}}$ becomes larger, and hence the local minimum of the total force curve at small $\tilde{r}_d$ becomes shallower. For that reason, the transition value of $\tilde{\xi}_{\mathrm{mag}}$ becomes larger (transition magnetic field becomes smaller), and the discontinuous jump in $\tilde{r}_d$ becomes smaller. Eventually, when $\tilde{r}_c\approx0.103$, the local minimum of the total force curve becomes an inflection point. In that case, the discontinuous jump in $\tilde{r}_d$ goes to zero, and $\tilde{r}_d$ changes continuously as a function of $\tilde{\xi}_{\mathrm{mag}}$. This evolution from a discontinuous jump to continuous behavior is analogous to the liquid-gas transition in classical thermodynamics.

The transition value of $\tilde{\xi}_{\mathrm{mag}}$ (or magnetic field) depends on the two dimensionless ratios $\tilde{r}_c=r_c/R$ and $\tilde{I}=I/K$. Because the core radius $r_c$ is much smaller than the droplet radius $R$ in most experiments, it is interesting to determine the behavior for small $\tilde{r}_c$. Figure~\ref{fig8:firstordertransition}(d) shows numerical solutions for several values of $\tilde{r}_c$, with fixed $\tilde{I}=1$. We can see that transition value of $\tilde{\xi}_{\mathrm{mag}}$ goes to zero (transition magnetic field goes to infinity) in the limit of $\tilde{r}_c\to0$. These numerical results can approximately be described by the form $\tilde{\xi}_{\mathrm{mag}}\sim 1/\log(1/\tilde{r}_c)$, as shown by the black dashed line. From the transition value of $\tilde{\xi}_{\mathrm{mag}}$, we can extract the transition value of the magnetic field using the relationship \begin{equation}\label{sec5:fieldprediction}
B=\frac{1}{R\tilde{\xi}_{\mathrm{mag}}}\sqrt{\frac{K\mu_0}{\chi_a}}\sim\frac{\log R}{R}.
\end{equation}
In the inset of Figure~\ref{fig8:firstordertransition}(d), we plot this prediction, assuming the parameters $I=K=10^{-11}$ N, $\chi_a = 10^{-6}$, $r_c=10$ nm, and $R=1$ to $100$~$\mu$m. The plot is in reasonable agreement with the experimental results reported in Ref.~\cite{ettinger2023magnetic}, considering the uncertainty in the parameters.

\section{Discussion and Conclusions}

In this article, we have applied the Peach-Koehler force approach to nematic liquid crystals in four different geometries, in order to determine their equilibrium configurations of disclination lines. This information is not directly available by solving the Euler-Lagrange equation for the director field, because the Euler-Lagrange equation assumes specific locations for the disclinations. In each of the examples, we follow the same general procedure: Determine the director field either analytically or numerically, find the total angular stress tensor, subtract off the angular stress of the corresponding free disclination to obtain the external angular stress acting on the disclination, and put that external angular stress tensor into the expression for the Peach-Koehler force. The equilibrium position of the disclination is then the position where the force vanishes.

In the first example, we consider a single straight twist disclination line between two infinite parallel substrates with uniform anchoring. The interaction between the disclination line and the two parallel substrates is derived both analytically and numerically using the Peach-Koehler force approach. This calculation identifies the equilibrium position of the disclination, and demonstrates the basic procedure for implementing the Peach-Koehler force approach. Second, we study a row of twist disclination lines between a patterned and a uniform substrate. In addition to the usual force repelling the disclinations from the substrates, the Peach-Koehler force approach also reveals a change in the force caused by the periodicity of the patterned substrate, which pushes the disclinations toward that substrate. Our theory predicts how the equilibrium positions of the disclinations depend on the ratio of the cell thickness to the patterned substrate periodicity. Third, we investigate two parallel $+1/2$ disclination lines inside a long cylindrical capillary tube with strong homeotropic anchoring. Our Peach-Koehler force calculation shows explicitly how the interaction between the disclinations and the cylindrical wall can be interpreted as the interaction between disclinations and their image defect lines. In addition, the analytical solution for the director field inside the 2D disk, derived from conformal mapping, also provides a good approximation for the nematic liquid crystal in a 3D spherical droplet in the next section.

In the fourth case, we have a radial hedgehog or disclination loop inside a sphere with strong homeotropic anchoring. This case is more complex than the previous examples, because the disclination is curved. The curved disclination experiences the Peach-Koehler force, due to self-interaction as well as repulsion from the spherical boundary. It also experiences the tension force, due to the energy cost of the defect core, and a new force arising from the spatial variation of the rotation vector $\hat{\boldsymbol{\Omega}}$ in the disclination loop. These three forces combine to give the equilibrium radius of the disclination loop, which depends on two dimensionless ratios---the line tension to the Frank elastic constant, and the disclination core radius to the droplet radius. Furthermore, we consider the effect of an applied magnetic field on the nematic droplet, and find how the loop radius changes as a function of magnetic field. This calculation shows that the loop may grow either continuously or discontinuously as the field is increased, depending on the same two dimensionless ratios, and it predicts the the field strength at the discontinuous transition. This problem of a radial hedgehog or disclination loop is a subject of current experimental interest~\cite{ettinger2023magnetic}, and the Peach-Koehler force approach may help to interpret further experiments on this system.

\bibliographystyle{tfnlm}
\bibliography{interactnlmsample}

\begin{thebibliography}{10}
\providecommand{\url}[1]{\normalfont{#1}}
\providecommand{\urlprefix}{Available from: }

\bibitem{brandenberger1994topological}
Brandenberger~RH. Topological defects and structure formation. Int J Mod Phys
  A. 1994;\hspace{0pt}9:2117--2189.

\bibitem{durrer2002cosmic}
Durrer~R, Kunz~M, Melchiorri~A. Cosmic structure formation with topological
  defects. Phys Rep. 2002;\hspace{0pt}364:1--81.

\bibitem{kosterlitz1973order}
Kosterlitz~JM, Thouless~DJ. Order, metastability and phase transitions in
  two-dimensional systems. J Phys C. 1973;\hspace{0pt}6:1181--1203.

\bibitem{agnolet1989kosterlitz}
Agnolet~G, McQueeney~DF, Reppy~JD. Kosterlitz-{Thouless} transition in helium
  films. Phys Rev B. 1989;\hspace{0pt}39:8934.

\bibitem{chaikin1995principles}
Chaikin~PM, Lubensky~TC. Principles of condensed matter physics. Cambridge;
  1995.

\bibitem{frank1950multiplication}
Frank~FC, Read~WT. Multiplication processes for slow moving dislocations. Phys
  Rev. 1950;\hspace{0pt}79:722.

\bibitem{halperin1978theory}
Halperin~BI, Nelson~DR. Theory of two-dimensional melting. Phys Rev Lett.
  1978;\hspace{0pt}41:121.

\bibitem{peng2016command}
Peng~C, Turiv~T, Guo~Y, et~al. Command of active matter by topological defects
  and patterns. Science. 2016;\hspace{0pt}354:882--885.

\bibitem{genkin2017topological}
Genkin~MM, Sokolov~A, Lavrentovich~OD, et~al. Topological defects in a living
  nematic ensnare swimming bacteria. Phys Rev X. 2017;\hspace{0pt}7:011029.

\bibitem{zhang2021spatiotemporal}
Zhang~R, Redford~SA, Ruijgrok~PV, et~al. Spatiotemporal control of liquid
  crystal structure and dynamics through activity patterning. Nat Mater.
  2021;\hspace{0pt}20:875--882.

\bibitem{shankar2022topological}
Shankar~S, Souslov~A, Bowick~MJ, et~al. Topological active matter. Nat Rev
  Phys. 2022;\hspace{0pt}4:380--398.

\bibitem{ardavseva2022topological}
Arda{\v{s}}eva~A, Doostmohammadi~A. Topological defects in biological matter.
  Nat Rev Phys. 2022;\hspace{0pt}4:354--356.

\bibitem{peach1950forces}
Peach~M, Koehler~JS. The forces exerted on dislocations and the stress fields
  produced by them. Phys Rev. 1950;\hspace{0pt}80:436--439.

\bibitem{anderson2017theory}
Anderson~PM, Hirth~JP, Lothe~J. Theory of dislocations. Cambridge; 2017.

\bibitem{de1995physics}
de~Gennes~PG, Prost~J. The physics of liquid crystals. Oxford; 1993.

\bibitem{kleman1983points}
Kl{\'e}man~M. Points, lines, and walls: In liquid crystals, magnetic systems,
  and various ordered media. Wiley; 1983.

\bibitem{eshelby1980force}
Eshelby~JD. The force on a disclination in a liquid crystal. Philos Mag A.
  1980;\hspace{0pt}42:359--367.

\bibitem{kawasaki1985gauge}
Kawasaki~K, Brand~HR. Gauge theory of continuous media with topological
  defects: Uniaxial nematic liquid crystals and superfluid $^4${He}. Ann Phys.
  1985;\hspace{0pt}160:420--440.

\bibitem{rey1990defect}
Rey~AD. Defect controlled dynamics of nematic liquids. Liq Cryst.
  1990;\hspace{0pt}7:315--334.

\bibitem{long2021geometry}
Long~C, Tang~X, Selinger~RLB, et~al. Geometry and mechanics of disclination
  lines in {3D} nematic liquid crystals. Soft Matter.
  2021;\hspace{0pt}17:2265--2278.

\bibitem{schimming2022singularity}
Schimming~CD, Vi{\~n}als~J. Singularity identification for the characterization
  of topology, geometry, and motion of nematic disclination lines. Soft Matter.
  2022;\hspace{0pt}18:2234--2244.

\bibitem{schimming2023kinematics}
Schimming~CD, Vi{\~n}als~J. Kinematics and dynamics of disclination lines in
  three-dimensional nematics. Proc R Soc A. 2023;\hspace{0pt}479:20230042.

\bibitem{long2022frank}
Long~C, Deutsch~MJ, Angelo~J, et~al. Frank-{R}ead mechanism in nematic liquid
  crystals; 2022. \urlprefix\url{https://arxiv.org/abs/2212.01316}.

\bibitem{wang2017artificial}
Wang~M, Li~Y, Yokoyama~H. Artificial web of disclination lines in nematic
  liquid crystals. Nat Commun. 2017;\hspace{0pt}8:388.

\bibitem{babakhanova2020dynamically}
Babakhanova~G, Golestani~YM, Baza~H, et~al. Dynamically morphing microchannels
  in liquid crystal elastomer coatings containing disclinations. J Appl Phys.
  2020;\hspace{0pt}128:184702.

\bibitem{allender1991determination}
Allender~DW, Crawford~GP, Doane~JW. Determination of the liquid-crystal surface
  elastic constant ${K}_{24}$. Phys Rev Lett. 1991;\hspace{0pt}67:1442--1445.

\bibitem{kralj1993stability}
Kralj~S, {\v{Z}}umer~S. The stability diagram of a nematic liquid crystal
  confined to a cylindrical cavity. Liq Cryst. 1993;\hspace{0pt}15:521--527.

\bibitem{kralj1995saddle}
Kralj~S, {\v{Z}}umer~S. Saddle-splay elasticity of nematic structures confined
  to a cylindrical capillary. Phys Rev E. 1995;\hspace{0pt}51:366--379.

\bibitem{burylov1997equilibrium}
Burylov~SV. Equilibrium configuration of a nematic liquid crystal confined to a
  cylindrical cavity. J Exp Theor Phys. 1997;\hspace{0pt}85:873--886.

\bibitem{shams2014theoretical}
Shams~A, Yao~X, Park~JO, et~al. Theoretical predictions of disclination loop
  growth for nematic liquid crystals under capillary confinement. Phys Rev E.
  2014;\hspace{0pt}90:042501.

\bibitem{de2007point}
De~Luca~G, Rey~AD. Point and ring defects in nematics under capillary
  confinement. J Chem Phys. 2007;\hspace{0pt}127:104902.

\bibitem{vafa2022defect}
Vafa~F, Zhang~GH, Nelson~DR. Defect absorption and emission for $p$-atic liquid
  crystals on cones. Phys Rev E. 2022;\hspace{0pt}106:024704.

\bibitem{kleman2006topological}
Kleman~M, Lavrentovich~OD. Topological point defects in nematic liquid
  crystals. Philos Mag. 2006;\hspace{0pt}86:4117--4137.

\bibitem{kralj1992freedericksz}
Kralj~S, {\v{Z}}umer~S. Fr{\'e}edericksz transitions in supra-$\mu$m nematic
  droplets. Phys Rev A. 1992;\hspace{0pt}45:2461--2470.

\bibitem{terentjev1995disclination}
Terentjev~EM. Disclination loops, standing alone and around solid particles, in
  nematic liquid crystals. Phys Rev E. 1995;\hspace{0pt}51:1330--1337.

\bibitem{fukuda2002stability}
Fukuda~J, Yokoyama~H. {Stability of a hyperbolic disclination ring in a nematic
  liquid crystal}. Phys Rev E. 2002;\hspace{0pt}66:012703.

\bibitem{wang2016experimental}
Wang~X, Kim~YK, Bukusoglu~E, et~al. Experimental insights into the
  nanostructure of the cores of topological defects in liquid crystals. Phys
  Rev Lett. 2016;\hspace{0pt}116:147801.

\bibitem{ettinger2023magnetic}
Ettinger~S, Slaughter~CG, Parra~SH, et~al. Magnetic-field-driven director
  configuration transitions in radial nematic liquid crystal droplets. Phys Rev
  E. 2023;\hspace{0pt}108:024704.

\end{thebibliography}

\end{document}